%% file: Derandomized.tex
\documentclass[11pt]{article}
\usepackage[utf8]{inputenc}
\usepackage[T1]{fontenc}
\usepackage[sc]{mathpazo}
\usepackage{microtype}
\usepackage{amsmath,xcolor}
\usepackage{amssymb}
\usepackage{amsthm}
\usepackage{bm,float}
\usepackage{dsfont}
\usepackage{authblk}
\usepackage{fullpage,caption,wrapfig}
\usepackage{comment}
\usepackage{mathtools}
\usepackage[shortlabels]{enumitem}
\usepackage{complexity}
\usepackage[backend=bibtex, style=alphabetic, backref=true,maxbibnames=99, url=false]{biblatex} 
\usepackage{bbm}
\usepackage{tcolorbox}

\usepackage{nag,tikz}
\usetikzlibrary{calc}
\usetikzlibrary{decorations.pathreplacing}
\usetikzlibrary{shapes, patterns, decorations, fit, intersections, arrows, automata, positioning}

\usepackage{algorithm,algpseudocode}
\usepackage{multicol}
\usepackage[scaled]{helvet} 
\usepackage{thmtools} 
\usepackage{hyperref}
\hypersetup{
    colorlinks=true,
    linkcolor=violet,
    filecolor=magenta,      
    urlcolor=cyan,
    citecolor=blue,
    pdffitwindow=true,
}\usepackage[capitalize, nameinlink]{cleveref}

\theoremstyle{plain}
\newtheorem{theorem}{Theorem}[section]
\newtheorem{lemma}[theorem]{Lemma}
\newtheorem{corollary}[theorem]{Corollary}
\newtheorem{proposition}[theorem]{Proposition}

\newtheorem{property}[theorem]{Property}

\newtheorem{definition}[theorem]{Definition}

\newtheorem{remark}[theorem]{Remark}

\newcommand{\old}[1]{}
\def\cR{{\cal R}}

\newcommand{\st}{\text{s.t.}}

\renewcommand{\R}{\ensuremath{\mathbb R}}

\renewcommand{\P}[1]{{\mathbb{P}}\left[#1\right]}
\renewcommand{\PP}[2]{{\mathbb{P}}_{#1}\left[#2\right]}

\newcommand{\I}[1]{ {\mathbb{I}}\left\{#1\right\} }

\renewcommand{\E}[1]{{\mathbb{E}}\left[#1\right]}
\renewcommand{\EE}[2]{{\mathbb{E}}_{#1}\left[#2\right]}

\renewcommand{\path}[2]{{ S_{#1}, \ldots, S_{#2} }}

\def\argmax{\textup{argmax}}

\def\b1{{\bf 1}}
\def\1{{\bf 1}}

\def\cB{{\cal B}}

\def\C{{\cal C}}

\def\eps{{\epsilon}}

\def\cU{{\cal U}}
\def\cI{{\cal I}}
\def\cH{{\cal H}}
\def\cT{{\cal T}}

\def\cA{{\cal A}}
\def\cN{{\cal N}}

\def\cE{{\cal E}}
\def\cost{c}
\def\cL{{\cal L}}

\def\R{\mathbb{R}}
\def\cR{{\mathcal{R}}}

\def\cC{{\cal C}}

\def\bbe{{\bf e}}
\def\bbf{{\bf f}}

\def\cO{{\cal C}}
\def\decrease{\beta}

\newtcolorbox{mybox}[3][]
{
  colframe = black,
  colback  = #2!10,  
  coltitle = white,  
  title    = {#3},
  #1,
}

\newcommand{\declareperson}[1]{\expandafter\newcommand\csname#1\endcsname[1]{\textcolor{orange}{#1: ##1}}}

\declareperson{Anna}
\declareperson{Shayan}
\declareperson{Nathan}
\hypersetup{ colorlinks=true, linkcolor=blue, filecolor=magenta, urlcolor=blue, }

\definecolor{arylideyellow}{rgb}{0.91, 0.84, 0.42}

\begin{document}

\title{A (Slightly) Improved Deterministic Approximation Algorithm for Metric TSP}
\author{Anna R. Karlin\thanks{\href{mailto:karlin@cs.washington.edu}{karlin@cs.washington.edu}. Research supported by Air Force Office of Scientific Research grant FA9550-20-1-0212 and NSF grant CCF-1813135.}}
\author{Nathan Klein\thanks{\href{mailto:nwklein@cs.washington.edu}{nwklein@cs.washington.edu}. Research supported in part by Air Force Office of Scientific Research grant FA9550-20-1-0212 and NSF grants DGE-1762114, CCF-1813135.}}
\author{Shayan Oveis Gharan\thanks{\href{mailto:shayan@cs.washington.edu}{shayan@cs.washington.edu}. Research supported by Air Force Office of Scientific Research grant FA9550-20-1-0212, NSF grants CCF-2203541,  and a Simons Investigator Award.}} 
\affil{University of Washington}


\maketitle
\begin{abstract}
We show that the max entropy algorithm can be derandomized (with respect to a particular objective function) to give a deterministic $3/2-\epsilon$ approximation algorithm for metric TSP for some $\epsilon > 10^{-36}$. 

To obtain our result, we apply the method of conditional expectation to an objective function constructed in prior work which was used to certify that the expected cost of the algorithm is at most $3/2-\epsilon$ times the cost of an optimal solution to the subtour elimination LP. The proof in this work involves showing that the expected value of this objective function can be computed in polynomial time (at all stages of the algorithm's execution).
\end{abstract}
\newpage

\input{introduction}

\input{preliminaries}

\input{degree-cut}

\input{extended-probabilistic}

\input{general-derand-overview}
\input{appendix}
\input{overview-laminar.tex}

\section{Acknowledgments} 

We would like to thank Andr{\'a}s Seb{\"o} for encouraging us to study this question and Martin N{\"a}gele for a helpful discussion on generating functions.

\printbibliography


\end{document}

%% file: introduction.tex
\section{Introduction}

One of the most fundamental problems in combinatorial optimization is the traveling salesperson problem (TSP), formalized as early as 1832 (c.f. \cite[Ch 1]{ABCC07}).
In an instance of  TSP we are given a set of $n$ cities $V$ along with their pairwise symmetric distances, $c:V\times V \to\R_{\geq 0}$. The goal is to find a Hamiltonian cycle of minimum cost. In the metric TSP problem, which we study here, the distances satisfy the triangle inequality. Therefore, the problem is equivalent to finding a closed Eulerian connected walk of minimum cost.

It is NP-hard to approximate TSP within a factor of $\frac{123}{122}$ \cite{KLS15}.  An algorithm of Christofides-Serdyukov~\cite{Chr76,Ser78} from four decades ago gives a $\frac32$-approximation for TSP.
Over the years there have been numerous attempts to improve the Christofides-Serdyukov algorithm and exciting progress has been made for various special cases of metric TSP, e.g., \cite{OSS11,MS11,Muc12,SV12,HNR21, KKO20, HN19, GLLM21}.
 Recently, ~\cite{KKO21a} gave the first improvement for the general case by demonstrating that the so-called ``max entropy" algorithm of the third author, Saberi, and Singh \cite{OSS11} gives a randomized $\frac{3}{2}-\epsilon$ approximation for some $\epsilon > 10^{-36}$.
 
 	The method introduced in \cite{KKO21a} exploits the optimum solution to the following linear programming relaxation of metric TSP studied by \cite{DFJ59,HK70,BG93}, also known as the subtour elimination LP:
\begin{equation}\label{eq:tsplp}
\begin{aligned}
	\min \quad& \sum_{u,v} x_{\{u,v\}} c(u,v)& \\
	\text{s.t.,} \quad &  \sum_{u} x_{\{u,v\}} = 2&\forall v\in V,\\
	& \sum_{u\in S, v\notin S} x_{\{u,v\}}\geq 2,&\forall S \subsetneq V, S\not= \emptyset\\
	& x_{\{u,v\}}\geq 0 &\forall u,v\in V.
\end{aligned}	
\end{equation} 
	
However, ~\cite{KKO21a} had two shortcomings. First, it did not show that the integrality gap of the subtour elimination polytope is bounded below $\frac{3}{2}$. Second, it was randomized, and the analysis in that work was by nature ``non-constructive" in the sense that it used the optimal solution; thus it was not clear how to to derandomize it using the method of conditional expectation. Other methods of derandomization seem at the moment out of reach and may require algorithmic breakthroughs. 
A followup work, \cite{KKO21b}, remedied the first shortcoming by showing an improved integrality gap. While it did not address the question of derandomization, a byproduct of that work is an analysis of the max entropy algorithm which is in principle polynomially-time computable as it avoids looking at OPT. The purpose of the present work is to show that this analysis can indeed be done in polynomial-time, from which the following can be deduced (remedying the second shortcoming of \cite{KKO21a}):

\begin{theorem}\label{thm:main}
	Let $x$ be a solution to LP \eqref{eq:tsplp} for a TSP instance. For some absolute constant $\epsilon > 10^{-36}$, there is a deterministic algorithm (in particular, a derandomized version of max entropy) which outputs a TSP tour with cost at most $\frac{3}{2}-\epsilon$ times the cost of $x$. 
\end{theorem} 

Thus, this work in some sense completes the exploratory program concerning whether the max entropy algorithm for TSP beats 3/2 (initiated by \cite{OSS11} in 2011), as now the above two weaknesses of \cite{KKO21a} have been addressed. Of course, much work remains in determining the true approximation factor of the algorithm; in this regard we are only at the tip of the iceburg.

Using the recent exciting work of Traub, Vygen, and Zenklusen reducing path TSP to TSP \cite{TVZ20} our theorem also implies that there is a deterministic $\frac{3}{2}-\epsilon$ approximation algorithm for path TSP.

\subsection{High level proof overview}
\label{sec:overview}


The high level strategy for derandomizing the max entropy algorithm is to use the method of conditional expectation on an objective function given by the analysis in \cite{KKO21b}.

The max entropy algorithm, similar to Christofides' algorithm, first selects a spanning tree and then adds a minimum cost matching on the odd vertices of the tree. While Christofides selects a minimum cost spanning tree, here the spanning tree is sampled from a distribution. In particular, after solving the natural LP relaxation for the problem to obtain a fractional solution $x$, a tree is sampled from the distribution $\mu$ which has maximal entropy subject to the constraint $\PP{T \sim \mu}{e \in T} = x_e$ for all $e \in E$ (with possibly some exponentially small error in these constraints). \cite{KKO21a,KKO21b} construct a so-called ``slack" vector which is used to show the expected cost of the matching (over the randomness of the trees) is at most $\frac{1}{2}-\epsilon$ times the cost of an optimal solution to the LP. Given a solution $x$ to LP \eqref{eq:tsplp} these works imply that there is a random vector $m$ as a function of the tree $T \sim \mu$ such that:
\begin{enumerate}
\item[(1)] The cost of the minimum cost matching on the odd vertices of tree $T$ is at most $c(m)$ (with probability 1), and
\item[(2)] $\EE{T \sim \mu}{c(m)} \le (\frac{1}{2}-\epsilon)c(x)$.
\end{enumerate}
Let $\cO = \EE{T \sim \mu}{c(T)+c(m)}$. This will be the objective function to which we will apply the method of conditional expectation. Since the expected cost of the tree $T$ is $c(x)$, as $\PP{T \sim \mu}{e \in T} = x_e$, by (2) $\cO$ is at most $(\frac{3}{2}-\epsilon)c(x)$. Since by (1) for a given tree $T$, $c(T)+c(m)$ is an upper bound on the cost of the output of the algorithm (with probability 1), this shows that the expected cost of the algorithm is bounded strictly below 3/2.

Ideally, one would like $\mu$ to have polynomial sized support. Then one could simply check the cost of the output of the algorithm on every tree in the support, and the above would guarantee that some tree gives a better-than-3/2 approximation. However, the max entropy distribution can have exponential sized support, and it's not clear how to find a similarly behaved distribution with polynomial sized support. 

Instead, let $\cT_{partial}$ be the family of all {\bf partial} settings of the edges of the graph to 0 or 1 where the edges set to 1 are acyclic. For $Set=\{X_{e_1},\dots,X_{e_i}\}\in \cT_{partial}$, and $1\leq j\leq i$, we use $X_{e_j}$ to indicate whether $e_j$ is set to $1$ or $0$.

The method of conditional expectations is then used 
as follows: Process the edges in an arbitrary order $e_1,\dots,e_m$ and for each edge $e_i$:
\begin{enumerate}
\item[(1)] Assume we inductively have chosen a valid assignment $Set\in \cT_{partial}$ to edges $e_1,\dots,e_{i-1}$. 
	\item[(2)] Let $Set^+=Set \cup \{X_{e_i}=1\}$. Compute $\cO^+= \EE{T \sim \mu}{c(T)+c(m) \mid Set^+}$. Similarly, let $Set^-=Set \cup \{X_{e_i}=0\}$ and compute $\cO^-=\EE{T \sim \mu}{c(T)+c(m) \mid Set^-}$.
	\item[(3)] Let $Set\gets Set^+$ or $Set\gets Set^-$ depending on which quantity is smaller.
\end{enumerate}
After a tree is obtained, add the minimum cost matching on the odd vertices of $T$. The resulting algorithm is shown in \cref{alg-derand} (see \cref{alg-derand-degree} for its instantiation in a simple case).

As $\cO \le (\frac{3}{2}-\epsilon)c(x)$, this algorithm succeeds with probability 1. We only need to show it can be made to run in polynomial time. Since we can compute the expected cost of the tree conditioned on $Set$ using linearity of expectation and the matrix tree theorem (\cref{sec:randalgoKKO}), it remains to show that $\EE{T \sim \mu}{c(m) | Set}$ can be computed deterministically and efficiently for any $Set\in \cT_{partial}$.

\vspace*{1mm}
{\bf Key Contributions.} The key contribution of this paper is to show how to do this computation efficiently, which is based on two observations: 
\begin{itemize}
	\item[(1)] The first is that the vector $m$ (whose cost upper bounds the cost of the minimum cost matching on the odd vertices of the tree) can be written as the (weighted) sum of indicators of events 
 that depend on the sampled tree $T$, and each of these events happens only when a constant number of (not necessarily disjoint) sets of edges have certain parities or certain sizes.
	\item[(2)] The second is that the probability of any such event can be deterministically computed in polynomial time by evaluating the generating polynomial of all spanning trees at certain points in $\mathbb{C}^{E}$, see \cref{lem:master-old}.
\end{itemize} 

\vspace*{1mm}
{\bf Structure of the paper.} After reviewing some preliminaries, in \cref{sec:computingprobabilities} we review the matrix tree theorem and show (as a warmup) how to compute the probability two (not necessarily disjoint) sets of edges both have an even number of edges in the sampled tree. In \cref{sec:degreecut}, we then give a complete description and proof of a deterministic algorithm for the special ``degree cut" case of TSP. Unlike the subsequent sections of the paper, \cref{sec:degreecut} is self-contained and thus directed towards readers looking for more high-level intuition or those not familiar with \cite{KKO21a,KKO21b}. In \cref{sec:probability} we show (2) from above and give the deterministic algorithm in the general case. The remainder  of the paper then involves proving (1) for the general definition of $m$ from \cite{KKO21a,KKO21b}.

\section{Preliminaries}
\subsection{Notation}

For a set of edges $A\subseteq E$ and (a tree) $T\subseteq E$, we write 
$\hypertarget{tar:AT}{A_T = |A \cap T|}.$
For a tree $T$, we will say a cut $S \subseteq V$ is \textit{odd in $T$} if $\delta(S)_T$ is odd and \textit{even in $T$} otherwise. If the tree is understood we will simply say even or odd. We use 
$\delta(S)=\{\{u,v\}\in E: |\{u,v\}\cap S|=1\}$ 
to denote the  set of edges that leave $S$, and $E(S) = \{\{u,v\}\in E: |\{u,v\}\cap S|=2\}$ to denote the set of edges inside of $S$.

For a set $A\subseteq E$ and a vector $x \in \R^{|E|}$ we write $ x(A):=\sum_{e\in A} x_e.$

\subsection{Randomized Algorithm of \cite{KKO21a}}\label{sec:randalgoKKO}


Let $x^0$ be an optimum solution of LP \eqref{eq:tsplp}. 
Without loss of generality we assume $x^0$ has an edge $e_0=\{u_0,v_0\}$ with $x^0_{e_0}=1, c(e_0)=0$.
(To justify this, consider the following process: given $x^0$, pick an arbitrary node, $u$, split it into two nodes $u_0,v_0$ and set $x_{\{u_0,v_0\}}=1, c(e_0)=0$ and assign half of every edge incident to $u$ to $u_0$ and the other half to $v_0$.) 

Let $E_0=E\cup\{e_0\}$ be the support of $x^0$ and let $x$ be $x^0$ restricted to $E$ and $G=(V,E)$. By \cref{fact:sptreepolytope} $x^0$ restricted to $E$ is in the spanning tree polytope  \eqref{eq:spanningtreelp} of $G$.
\hypertarget{tar:G=(V,E,x)}{We write $G=(V,E,x)$ to denote the (undirected) graph $G$ together with special vertices $u_0,v_0$ and the weight function $x:E\to\R_{\geq 0}$. Similarly, let $G_0 = (V,E_0,x^0)$ and let $G_{/ e_0} = G_0/\{e_0\}$, i.e. $G_{/ e_0}$ is the graph $G_0$ with the edge $e_0$ contracted}. 

\begin{definition}\label{defn:lambdaUniform}
For a vector $\lambda:E\to\R_{\geq 0}$, a $\lambda$-uniform distribution $\mu_\lambda$ over spanning trees of $G=(V,E)$ is a distribution where for every spanning tree $T\subseteq E$, $\PP{\mu_\lambda}{T}=\frac{\prod_{e\in T} \lambda_e}{\sum_{T'} \prod_{e\in T'} \lambda_e}$.	
\end{definition}


\begin{theorem}[\cite{AGMOS10}]
\label{thm:maxentropycomp}
Let $z$ be a point in the spanning tree polytope (see \eqref{eq:spanningtreelp}) of a graph $ G=(V, E)$.
For any $\eps>0$, a vector $\lambda:E\to\R_{\geq 0}$ can be found such that the corresponding $\lambda$-uniform spanning tree distribution, $\mu_\lambda$, satisfies
$$
\sum_{T\in {\cal T}: T \ni e} \PP{\mu_\lambda}{T}  \leq (1+\varepsilon)z_e,\hspace{3ex}\forall e\in E,$$
i.e., the marginals are approximately preserved.  In the above ${\cal T}$ is the set of all spanning trees of $(V,E)$. The algorithm is deterministic and running
time is polynomial in $n=|V|$, $- \log \min_{e\in E} z_e$ and $\log(1/\eps)$.
\end{theorem}

\cite{KKO21b} showed that	the following (randomized) max entropy algorithm  
has expected cost of the output is at most $(\frac{3}{2}-\epsilon)c(x)$.


\hypertarget{tar:alg}{\begin{algorithm}[h]
\begin{algorithmic}
	\State Find an optimum solution $x^0$ of  \cref{eq:tsplp}, and let $e_0=\{u_0,v_0\}$ be an edge with $x^0_{e_0}=1,c(e_0)=0$.
	\State Let $E_0=E\cup \{e_0\}$ be the support of $x^0$ and $x$ be $x^0$ restricted to $E$ and $G=(V,E)$.
	\State Find a vector $\lambda:E\to\R_{\geq 0}$ such that for any $e\in E$, $\PP{T \sim \mu_\lambda}{e \in T}=x_e(1\pm 2^{-n})$.
	\State Sample a tree $T\sim\mu_\lambda$.
	\State Let $M$ be the minimum cost matching on odd degree vertices of $T$.
	\State Output $T \cup M$.
\end{algorithmic}
\caption{(Randomized) Max Entropy Algorithm for TSP}\label{alg:tsp}
\end{algorithm}}

%% file: preliminaries.tex
\subsection{Polyhedral background}
For any graph $G=(V,E)$,
Edmonds \cite{Edm70} gave the following description for the convex hull of spanning trees of a graph $G=(V,E)$, known as the {\em spanning tree polytope}.
\begin{equation}
\begin{aligned}
 z(E) = |V|-1, \quad z(E(S)) \leq |S|-1\quad    \forall S\subseteq V, \quad z_e \geq 0  \quad \forall e\in E.
\end{aligned}
\label{eq:spanningtreelp}
\end{equation}
Edmonds \cite{Edm70} proved that the extreme point solutions of this polytope are the characteristic vectors of the spanning trees of $G$.

\begin{lemma}[{\cite[Fact 2.1]{KKO21a}}] \label{fact:sptreepolytope}
Let $x^0$ be a feasible solution of \eqref{eq:tsplp} such that $x^0_{e_0}=1$ with support $E_0=E\cup \{e_0\}$. 
Let $x$ be $x^0$ restricted to $E$; then $x$ is in the spanning tree polytope of $G=(V,E)$. 
\end{lemma}

Since $c(e_0)=0$, the following fact is immediate.
\begin{lemma} \label{fact:expcostT}Let $G=(V,E,x)$ where  $x$ is in the spanning tree polytope. If $\mu$ is any distribution of spanning trees with marginals $x$ then $\EE{T\sim\mu}{c(T \cup e_{0})}=c(x)$.
 \end{lemma}
 
  To bound the cost of the min-cost matching on the set $O(T)$ of odd degree vertices of the tree $T$, we use the following characterization of the $O(T)$-join polyhedron 
  due to Edmonds and Johnson \cite{EJ73}.
\begin{proposition}
\label{prop:tjoin}
For any graph $G=(V,E)$, cost function $c: E \to \R_+$, and a set $O\subseteq V$ with an even number of vertices,  the minimum weight of an $O$-join equals the optimum value of the following integral linear program.
\begin{equation}
\begin{aligned}
&\min \hspace{4ex}  \cost(y) \quad \st\\
&   y(\delta(S)) \geq 1 \quad \hspace{0.2cm} S \subseteq V, |S\cap  O| \text{ odd} \quad 
 y_e \geq 0 \hspace{0.2cm} \forall e\in E
\end{aligned}
\label{eq:tjoinlp}
\end{equation}
\end{proposition}

\section{Computing probabilities}\label{sec:computingprobabilities}

The deterministic algorithm depends on the computation of various probabilities and conditional expectations. In this section (and additionally later in \cref{sec:probability}), we  show to do these calculations efficiently.

\subsection{Notation}

 Let $\cB_E$ be the set of all probability measures on the Boolean algebra $2^{|E|}$. Let $\mu\in\cB_E$. The generating polynomial $g_\mu: \R[\{z_{e}\}_{e\in E}]$ of $\mu$ is defined as follows:
$$ g_\mu(z):=\sum_S \mu(S) \prod_{e\in S} z_e.$$

\subsection{Matrix tree theorem}

Let $G=(V,E)$ with $|V|=n$. For $e=(u,v)$ we let $L_e = (\1_u - \1_v)(\1_u - \1_v)^T$ be the Laplacian of $e$. Recall Kirchhoff's matrix tree theorem: 


\begin{theorem}[Matrix tree theorem]\label{thm:matrixtree}
	For a graph $G=(V,E)$ let $g_\cT \in \R[z_{e_1},\dots,z_{e_m}] = \sum_{T \in \cT} z^T$ be the generating polynomial of the spanning trees of $G$.
	
	Then, we have
	$$g_\cT(\{z_e\}_{e\in E})  = \frac{1}{n}\det(\sum_{e \in E} z_eL_e + 11^T/n).$$
  
\end{theorem}


Given a vector $\lambda \in \R^{|E|}$ and a set $S \subseteq E$, let $\lambda^S := \prod_{i \in S} \lambda_i$. 
Recall that the  $\lambda$-uniform distribution $\mu_\lambda$ is the probability distribution over spanning trees where the probability of every tree $T$ is $\lambda^T$.
Then the generating polynomial of $\mu_\lambda$ is
$$ g_{\mu_\lambda}(z)=\sum_{T \in \cT} \lambda^Tz^T = g_{{\cal T}}(\{\lambda_e z_e\}_{e\in E}) = \frac1n\det\left(\sum_{e\in E} z_e\lambda_eL_e +11^T/n\right)$$
and can be evaluated at any $z \in  \mathbb{C}^{E}$ efficiently using a determinant computation.

Thus we can compute $\PP{T \sim \mu}{e \in T}$ by computing the sum of the probabilities of trees in the graph $G/\{e\}$, i.e. the graph with $e$ contracted, as follows: 
$$\PP{T \sim \mu}{e \in T} = 1-\PP{T\sim\mu}{e\notin T} =1-\sum_{T \in \cT:e\notin T} \lambda^T$$
where to compute the sum in the RHS we evaluate $g_{\mu_\lambda}$ at $z_e=0, z_f=1$ for all $f\neq e$. Thus, 
\begin{lemma}\label{fact:marginals}
Given a $\lambda$-uniform distribution $\mu_\lambda$ over spanning trees, for every edge $e$, we can compute $\PP{T \sim \mu_\lambda}{e \in T}$ in polynomial time. 
\end{lemma}

Given some $Set \in \cT_{partial}$, 
we contract  each edge $e$ with $X_e = 1$ in $Set$ and delete each edge $e$ with $X_e = 0$ in $Set$. Let $G'$ be the resulting graph with $n'$ vertices, with corresponding $\lambda'_e\propto \lambda_e$ for all $e\in G'$ normalized such that $\sum_{T'\in G'}{\lambda'}^{T}=1$.

\begin{remark}
A vector $\lambda \in \R^{|E|}$
is easily normalized by setting $\lambda_e'  = \lambda_e/\left(\sum_T \lambda^T\right)^{1/n-1}$, i.e.,
 $\lambda_e'  = \lambda_e/g_{{\cal T}}(\{\lambda_e\}_{e\in E})^{1/n-1}$. 
Thus at the cost of another application of the matrix-tree theorem, we assume without loss of generality that we are always dealing with 
$\lambda$ values that are normalized.
\end{remark}

Putting the previous facts together, we obtain
\begin{lemma}\label{fact:compute-new-lambda'}
	Given a $\lambda$-uniform distribution $\mu_\lambda$ and some $Set \in \cT_{partial}$, we can compute a vector $\lambda'$ such that $\mu_{\lambda'} = \mu_{\lambda \mid Set}$.
\end{lemma}



\subsection{Computing parities in a simple case}

\begin{lemma}
Let $A, B \subseteq E$ and $\mu_\lambda$ be a $\lambda$-uniform distribution over spanning trees. Then, we can compute $\PP{T \sim \mu_\lambda}{A_T, B_T \text{ even}}$ in polynomial time. 
\end{lemma}
\begin{proof}
First observe that
$$\I{A_T, B_T \text{ even}} = \frac{1}{4}(1 +  (-1)^{A_T} + (-1)^{B_T} + (-1)^{((A \smallsetminus B) \cup (B \smallsetminus A))_T})$$
One can easily check that if $A_T$ and $B_T$ are even, this is 1, and otherwise it is 0. 
	
To compute $\PP{T \sim \mu_\lambda}{A \text{ and } B \text{ even in } T}$ it is enough to compute the expected value of this indicator. By linearity of expectation it is therefore enough to compute the expectation of $(-1)^{F_T}$ for any set $F \subseteq E$. We can do this using \cref{thm:matrixtree}. Setting $z^F_e = -1$ if $e \in F$ and $z^F_e=+1$ otherwise, we exactly have:
$$g_{\mu_\lambda}(z^F) = \sum_{T \in \cT} (-1)^{F_T} \lambda^T = \EE{T \sim \mu_\lambda}{(-1)^{F_T}}.$$
The lemma follows.
\end{proof}
\begin{remark}
We can use the same approach to compute $\PP{T\sim\mu_\lambda}{A_T\text{ odd, } B_T \text{ even}}$ or the probability that both are odd. All we need to do is to multiply $(-1)^{A_T}$ with a $-1$ if $A_T$ needs to be odd (and similarly for $B_T$), and $(-1)^{((A\smallsetminus B)\cup (B\smallsetminus A))_T}$ with a $-1$ if we are looking for different parities in $A_T,B_T$. 
\end{remark}

Given some $Set \in \cT_{partial}$, by \cref{fact:compute-new-lambda'} we can compute $\mu_{\lambda'} = \mu_{\lambda \mid Set}$.  Applying the above lemma to $\mu_{\lambda'}$, it follows (after appropriately updating the parities to account for edges set to 1 in $Set$):
\begin{corollary}\label{cor:prob-degcutcase}
Let $A, B \subseteq E$. We can compute $\PP{T \sim \mu}{A \text{ and } B \text{ even in } T \mid Set}$ in polynomial time. 
\end{corollary}

%% file: degree-cut.tex
\section{A deterministic algorithm in the degree cut case}
\label{sec:degreecut}


As a warmup, in this section we show how to implement the deterministic algorithm for the so-called {\em``degree cut case,"} i.e., when for every set of vertices $S$ with $2 \le |S| \le n-2$ we have $x(\delta(S)) \ge 2+\eta$ for some absolute constant $\eta > 0$. See \cref{alg-derand-degree}.

\begin{algorithm}[h]
\caption{A Deterministic Approximation Algorithm for Metric TSP in the Degree Cut Case}
\begin{algorithmic}[1]
\State Given a solution $x^0$ of the LP \eqref{eq:tsplp}, with an edge $e_0$ with $x_{e_0}=1$.
\State Let $G$ be the support graph of $x$.
\State Find a vector $\lambda:E\to\R_{\geq 0}$ such that for any $e\in E$, $\PP{T\sim\mu_\lambda}{e\in T} = x_e(1\pm 2^{-n})$ (see \cref{sec:randalgoKKO}).
\State Initialize $Set := \emptyset$ 
\While {there exists $e\neq e_0$ not set in $Set$}
\State Let $Set^+  := Set \cup \{X_e = 1\}$ and let   $Set^-  := Set \cup \{X_e = 0\}$;
\If{\hyperlink{computing-deg}{$\EE{T\sim\mu_\lambda}{c(T)+c(m) ~|~ Set^+} \le \EE{T\sim\mu_\lambda}{c(T)+c(m) ~|~ Set^-}$} ($m$ from \cref{def:matching-deg-cut})}
\State $Set := Set^+$; 
\Else 
\State $Set := Set^-$; 
\EndIf
\EndWhile
\State Return $T=\{e: X_e=1 \text{ in Set}\}$ together with min cost  matching on odd degree vertices of $T$. 
\end{algorithmic}
\label{alg-derand-degree}
\end{algorithm}


{\bf Construction of the matching vector.} We describe a simple construction for the matching vector $m: \cT \to \R^{|E|}$ for the degree cut case. It will ensure that for a tree $T$, $m$ is in the $O(T)$-Join polyhedron where $O(T)$ is the set of odd vertices of $T$ (we emphasize that $m$ is a function of $T$). Therefore, $c(m)$ is an upper bound on the cost of the minimum cost matching on the odd vertices of $T$ as desired.

Let $p = 2 \cdot 10^{-10}$ (note that we have not optimized this constant and in the degree cut case it can be greatly improved). We say that an edge $e=(u,v)$ is \textbf{good} if\\ $\PP{T \sim \mu}{u,v \text{ both even in }T} \ge p$, where we say a vertex $v$ is even in a tree $T$ if $\delta(v)_T$ is even. The vector $m$ will consist of the convex combination of two feasible points in the $O(T)$-Join polyhedron, $g$ and $b$ (where $g$ is for ``good" edges and $b$ is for ``bad" edges). 

For a tree $T$ and an edge $e=(u,v)$ we let:
$$g_e = \begin{cases}	
	\frac{1}{2+\eta}x_e & \text{If $u$ and $v$ are both even in $T$} \\
		\frac{1}{2}x_e & \text{Otherwise} \\
\end{cases}
 $$
 
 \begin{lemma}
 $g$ is in the $O(T)$-Join polyhedron. 	
 \end{lemma}
\begin{proof}
First, consider any cut consisting of a single vertex $v$ (or its complement). If $v$ is odd, we need to ensure that $g(\delta(v)) \ge 1$. If $v$ is odd, then $g_e = x_e/2$ for all $e \in \delta(v)$, so this follows from the fact that $x(\delta(v)) = 2$. 	

Now consider any cut $S$ with $2 \le |S| \le n-2$. We now argue that $g(\delta(S)) \ge 1$ with probability 1. This follows from the fact that:
$$g(\delta(S)) \ge \frac{1}{2+\eta} x(\delta(S)) \ge \frac{1}{2+\eta}(2+\eta) = 1,$$
where we use that every cut $S$ with $2 \le |S| \le n-2$ has $x(\delta(S)) \ge 2+\eta$.
\end{proof}

 We now design our second vector $b$. For a tree $T$ and an edge $e=(u,v)$ we let:
$$b_e = \begin{cases} \frac{1+\eta}{2+\eta}x_e & \text{If $e$ is good} \\ \frac{1}{2+\eta}x_e & \text{If $e$ is bad} \end{cases}$$
 
We will crucially use the following:
\begin{corollary}[Corollary of Theorem 5.14 from \cite{KKO21a}]
	Let $v$ be a vertex. Then, if $G_v$ is the set of good edges adjacent to $v$, $x(G_v) \ge 1$. 
\end{corollary}

In \cite{KKO21a}, it is shown that if $x_e$ is bounded away from $1/2$, then $e$ is a good edge. Furthermore, for any two edges $e$ and $f$ adjacent to $v$ with $x_e \approx x_f \approx 1/2$, at least one is good. So, $v$ can have only one bad edge which has fraction about $1/2$, giving the above corollary (therefore it is even true that $x(G_v) \ge  3/2 - \gamma$ for some small $\gamma > 0$). 

Given this, we can show the following:
\begin{lemma}
	$b$ is in the $O(T)$-Join polyhedron. 
\end{lemma}
\begin{proof}
For any non-vertex cut, similar to above, the $O(T)$-Join constraint is easily satisfied. For a vertex cut $v$, we use that by the above theorem the $x$ weight of the set of good edges adjacent to $v$ is at least 1. Therefore, $b(v) \ge \frac{1+\eta}{2+\eta} + \frac{1}{2+\eta} = 1$.
\end{proof}

\begin{definition}[Matching vector $m$ in the degree cut case]\label{def:matching-deg-cut}
Let 
	$m = \alpha b + (1-\alpha)g,$ 
 for some $0 < \alpha < 1$ we choose in the next subsection. Since $b$ and $g$ are both in the $O(T)$-Join polyhedron, so is $m$.
\end{definition}

 \begin{lemma}
 For any good edge $e$, $\E{g_e} \le (\frac{1}{2} - \frac{\eta p}{4+2\eta})x_e$.	
 \end{lemma}
 \begin{proof}
 Let $p_e = \PP{T \sim \mu}{u,v \text{ even}}$. We can compute:
 	$$\E{g_e} = \left(\frac{p_e}{2+\eta} + \frac{1-p_e}{2}\right)x_e \le \left(\frac{p}{2+\eta} + \frac{1-p}{2}\right)x_e = \left(\frac{1}{2} - \frac{\eta p}{4+2\eta} \right)x_e,$$
 	as desired.
 \end{proof}

Therefore, for any good edge $e$,
$$\E{m_e} \le \left(\alpha \left(\frac{1+\eta}{2+\eta}\right) + (1-\alpha)\left(\frac{1}{2} - \frac{\eta p}{4+2\eta}\right)\right)x_e$$
For any bad edge $e$, we have
$$\E{m_e} \le  \left(\frac{\alpha}{2+\eta} + \frac{1-\alpha}{2}\right)x_e$$
To make the two equal, we set $\alpha = \frac{p}{2+p}$. Therefore, 
$$\E{m_e} \le \left(\frac{p/(2+p)}{2+\eta} + \frac{1-p/(2+p)}{2}\right)x_e < \left(\frac{1}{2} - \frac{p\eta}{9}\right)x_e$$ for all edges $e$. Since $\eta,p$ are absolute constants, this is at most $(\frac{1}{2}-\epsilon)x_e$ for some absolute constant $\epsilon > 0$. Therefore the randomized algorithm has expected cost at most $(\frac{3}{2}-\epsilon)c(x)$, which is enough to prove that \cref{alg-derand-degree} deterministically finds a tree plus a matching whose cost is at most $(\frac{3}{2}-\epsilon)c(x)$. Thus the only remaining question is the computational complexity of \cref{alg-derand-degree}, which we address now.

\hypertarget{computing-deg}{\paragraph{\bf Computing $\E{c(T) + c(m) \mid Set}$.}}
Now that we have explained the construction of $m$, we observe that there is a simple deterministic algorithm to compute $\E{c(T) + c(m) \mid Set}$ in polynomial time.

First, compute $\E{c(T) \mid Set}$. By linearity of expectation it is enough to compute $\P{e \in T \mid Set}$ for all $e \in E$. To do this, we first apply \cref{fact:compute-new-lambda'} to find $\lambda'$ such that $\mu_{\lambda'} = \mu_{\lambda \mid Set}$ and then apply \cref{fact:marginals}.

Now to compute $\E{c(m) \mid Set}$, it suffices to compute $\E{m_e \mid Set}$ for any \hyperlink{Tpartial}{$Set \in \cT_{partial}$}, $\P{e \in T \mid Set}$ and any $e=(u,v)$. Given the definition of $m$, the only event depending on the tree is the event $\P{u,v \text{ even} \mid Set}$. This can be computed with \cref{cor:prob-degcutcase}.

%% file: extended-probabilistic.tex
\section{General Case}\label{sec:probability}

\begin{algorithm}[h]
\caption{A Deterministic Approximation Algorithm for Metric TSP}
\begin{algorithmic}[1]
\State Given a solution $x^0$ of the LP \eqref{eq:tsplp}, with an edge $e_0$ with $x_{e_0}=1$.
\State Let $G$ be the support graph of $x$.
\State Find a vector $\lambda:E\to\R_{\geq 0}$ such that for any $e\in E$, $\PP{T\sim\mu_\lambda}{e\in T} = x_e(1\pm 2^{-n})$
\State Perform Preprocessing Steps \hyperlink{preprocessing1}{1}, \hyperlink{preprocessing2}{2}, \hyperlink{preprocessing3}{3}, \hyperlink{preprocessing4}{4}, \hyperlink{preprocessing5}{5}, and \hyperlink{preprocessing6}{6}
\State Initialize $Set := \emptyset$.
\While {there exists $e\neq e_0$ not set in $Set$}
\State Let $Set^+  := Set \cup \{X_e = 1\}$ and let   $Set^-  := Set \cup \{X_e = 0\}$;
\State Compute $S^+ = \EE{T\sim\mu_\lambda}{c(T) ~|~ Set^+}+\sum_{e \in E}\hyperlink{csstar}{\mathbb{E}_{c(s^*)}(e,Set^+)} + \hyperlink{cs}{\mathbb{E}_{c(s)}(e,Set^+)}$.
\State Compute $S^- = \EE{T\sim\mu_\lambda}{c(T) ~|~ Set^-}+\sum_{e \in E}\hyperlink{csstar}{\mathbb{E}_{c(s^*)}(e,Set^-)} + \hyperlink{cs}{\mathbb{E}_{c(s)}(e,Set^-)}$.
\State If $S^+ \le S^-$, let $Set := Set^+$. Otherwise let $Set := Set^-$.
\EndWhile
\State Return $T=\{e: X_e=1 \text{ in Set}\}$ together with min cost  matching on odd degree vertices of $T$. 
\end{algorithmic}
\label{alg-derand}
\end{algorithm}

The matching vector $m$ in the general case, \cite[Thm 6.1]{KKO21b}, can be written as $s+s^*+\frac{1}{2}x$ where $s,s^*$ are  functions of the tree $T\sim \mu_{\lambda}$ and some independent Bernoullis $\cB$.
Roughly speaking, the (slack) vector $s^*:E\to\R_{\geq 0}$ takes care of matching constraints for near minimum cuts that are crossed and the (slack) vector $s:E\to\R$ takes care of the constraints corresponding to cuts which are not crossed.
Most importantly, the guarantee is that for a fixed tree $T$ the expectation of $c(s) + c(s^*) + \frac{1}{2}c(x)$ over the Bernoullis is at least $c(M)$ where $M$ is the minimum cost matching on the odd vertices of $T$. Furthermore, $\E{c(s) + c(s^*)}\leq -\eps c(x)$ which is the necessary bound to begin applying the method of conditional expectation in \cref{alg-derand}.

\begin{remark}
	The definitions of $s$ and $s^*$,  
 the proof that $\E{c(s) + c(s^*)} \le -\epsilon c(x)$, and the proof that $x/2+\E{s + s^* \mid T}$ is in the $O(T)$-join polyhedron come from \cite{KKO21a,KKO21b}. Here, we will review how to construct the random slack vectors $s,s^*$ for a given spanning tree $T$ and then explain how to efficiently compute $\E{c(s) + c(s^*) \mid Set}$ deterministically for any $Set \in \cT_{partial}$. 
	
	 Unfortunately, a reader who has not read \cite{KKO21a,KKO21b} may not be able to understand the motivation behind the details of the construction of $s,s^*$. However, \cref{sec:preprocessing-sstar} and \cref{sec:preprocessing-s} are self-contained in the sense that a reader should be able to verify that $\E{c(s) + c(s^*) \mid Set}$ can be computed efficiently and deterministically.
\end{remark}

Our theorem boils down to showing the following two lemmas:
\begin{lemma}\label{lem:compute-s}
For any  $Set \in \cT_{partial}$, there is a polynomial time deterministic algorithm that computes:
\begin{enumerate}[(1)]
\item  $\EE{T \sim \mu_\lambda}{c(s^*) \mid Set}$ (shown in $\hyperlink{csstar}{\mathbb{E}_{c(s^*)}(e,Set)}$)
\item $\EE{T \sim \mu_\lambda}{c(s) \mid Set}$ (shown in $\hyperlink{cs}{\mathbb{E}_{c(s)}(e,Set)}$)
\end{enumerate}
 
\end{lemma}

The crux of proving the above lemma is to show that for a given edge $e$ and any $Set$, each of $\E{s^*_e \mid Set}$ and $\E{s_e \mid Set}$ can be written as the (weighted) sum of indicators of events that depend on the sampled tree $T$, and each of these events happens only when a constant number of (not necessarily disjoint) sets of edges have certain parities or certain sizes.
Technically speaking, these weighted sums are  non-trivial for some of the events defined in \cite{KKO21a,KKO21b}. 
Given that, the following is enough to prove \cref{lem:compute-s}, as it gives a deterministic algorithm to compute the probability that a collection of (not necessarily disjoint) sets of edges have certain parities or certain sizes.

(1) of \cref{lem:compute-s} is proved in \cref{sec:preprocessing-sstar}, and (2) in \cref{sec:preprocessing-s}. The algorithm for each part requires a series of preprocessing steps and function definitions that we have marked with gray boxes. In each section, the final procedure to calculate the expected cost of the slack vector is given in a yellow box at the end of the corresponding section.

\begin{lemma}\label{lem:master-old}
Given a probability distribution $\mu:2^{[n]}\to\R_{\geq 0}$ and an oracle $O$ that can evaluate $g_\mu(z_1,\dots,z_n)$ at any $z_1,\dots,z_n\in\mathbb{C}$.
Let $E_1,\dots,E_k$ 	be a collection of (not necessarily disjoint) subsets of $[n]$ and $(\sigma_1,\dots,\sigma_k)\in \mathbb{F}_{m_1}\times \dots\times \mathbb{F}_{m_k}$. Then, we can compute,
$$ \PP{T \sim \mu}{(E_i)_T = \sigma_i (\text{mod }m_i), \forall 1\leq i\leq k}. $$ 
in $N:=m_1\dots m_k$-many calls to the oracle.\footnote{Note that since we are dealing with irrational numbers, we will not be able to compute this probability exactly. However by doing all calculations with $poly(n,N)$ bits of precision we can ensure our estimate has exponentially small error which will suffice to get the bounds we need later.}
\end{lemma}
\begin{proof}
For each of the sets $E_i$, define a variable $x_i$, and substitute for $z_e\gets \prod_{j} x_j^{\I{e \in E_j}}$ into the polynomial $g_\mu$ and call the resulting polynomial $g$.
Then $$g(x_1, \ldots, x_k) = \sum_{S \in supp(\mu)} \P{S} \prod_{i=1}^k x_i^{(E_i)_S}  
$$
Where recall $(E_i)_S = |E_i \cap S|$. Now, let $\omega_i := e^{\frac{2\pi\sqrt{-1} }{m_i}}$. 
We claim that
$$ \frac{1}{m_1\cdots m_k}\sum_{(e_1,\dots,e_k)\in \mathbb{F}_{r_1} \times \dots \times \mathbb{F}_{r_k}} \prod_{i=1}^k \omega_i^{-e_i\sigma_i} g(\omega_1^{e_1},\dots,\omega_k^{e_k}) = \PP{S \sim \mu}{(E_i)_S \equiv \sigma_i \bmod m_i, \forall 1\leq i\leq k}$$
So the algorithm only needs to call the oracle $N$ many times to compute the sum in the LHS.

To see this identity, notice that we can write the LHS as
\begin{align*} 
\frac{1}{m_1\cdots m_k} 
\sum_{(e_1,\dots,e_k)\in \mathbb{F}_{m_1} \times \dots \times \mathbb{F}_{m_k}} 
\sum_{S\in supp(\mu)} \P{S}
\prod_{i=1}^k \omega_i^{-e_i\sigma_i+e_i(E_i)_S} &=\sum_{S\in supp(\mu)}\P{S}\prod_{i=1}^k \left(\frac{1}{m_i} \sum_{e_i\in\mathbb{F}_{m_i}} \omega_i^{((E_i)_S-\sigma_i)e_i}\right)\\
&= \sum_{S\in supp(\mu)}\P{S}\prod_{i=1}^k\I{(E_i)_S-\sigma_i\equiv 0\bmod \sigma_i}
\end{align*}
where the last equality uses that $\omega_i$ is the $m_i$'th root of unity. 
The RHS is exactly equal to the probability that $(E_i)_S\equiv \sigma_i\bmod m_i$ for all $i$.
\end{proof}

\begin{remark}
	When we apply this lemma in this paper, we will always let $k$ be a constant and $m_i \le |V|$ for all $i$. Thus, it will always use a polynomial number of calls to an oracle evaluating the generating polynomial of a spanning tree distribution $\mu_\lambda$. By \cref{thm:matrixtree}, for any $z \in \mathbb{C}^{|E|}$:
	$$g_{\mu_\lambda}(\{z_e\}_{e \in E}) = \frac{1}{n}\det(\sum_{e \in E} \lambda_e z_e L_e + 11^T/n),$$   
	which can be computed in polynomial time.
\end{remark}

\begin{corollary}\label{lem:master}
Let $\mu_\lambda$ be a $\lambda$-uniform spanning tree distribution and let \hyperlink{Tpartial}{$Set \in \cT_{partial}$.} Then, let $E_1,\dots,E_k$ 	be a collection of (not necessarily disjoint) subsets of $[n]$ and $(\sigma_1,\dots,\sigma_k)\in \mathbb{F}_{m_1}\times \dots\times \mathbb{F}_{m_k}$. Then, we can compute,
$$ \PP{T \sim \mu_\lambda}{(E_i)_T = \sigma_i \pmod{m_i}, \forall 1\leq i\leq k \mid Set}. $$ 
in $N:=m_1\dots m_k$-many calls to the oracle.
\end{corollary}
\begin{proof}
Construct a new graph $G'$ by contracting all edges with $X_e=1$ in $Set$ and deleting all edges with $X_e=0$. We then update all $\sigma_i$ by subtracting the number of edges that are set to 1 in $E_i$ by $Set$. Then we apply \cref{lem:master-old} to the $\lambda$-uniform spanning tree distribution over $G'$ with the updated $\vec{\sigma}$ and the same $\vec{m}$.
\end{proof}

%% file: appendix.tex
\section{Computation for $c(s^*)$}\label{sec:preprocessing-sstar}

We interleave the definitions of $s$ and $s^*$ with our method of computing the expected value of these vectors. While the definitions of $s,s^*$ are essentially copied from \cite{KKO21a} and \cite{KKO21b}, in some places we modify the notation and make the construction slightly more algorithmic to improve the presentation. To differentiate these two we put computations in boxes. 

\hypertarget{tar:crossing}{For two sets $A,B\subseteq V$, we say $A$ {\em crosses} $B$ if all of the following sets are non-empty:
$$ A\cap B, A\smallsetminus B, B\smallsetminus A, \overline{A\cup B}.$$}

\begin{definition}[Near Min Cut]\hypertarget{tar:nearmincut}{For $G=(V,E,x)$, we say a cut $S\subseteq V$ is an {\em $\eta$-near min cut} if $x(\delta(S))< 2+\eta$.\footnote{Note this differs slightly from the notation in \cite{Ben95, BG08} in which an $\eta$ near min cut is said to be within a $1+\eta$ factor of the edge connectivity of the graph.}}
\end{definition}



\subsection{Polygon representation preprocessing}

\begin{definition}[Connected Component of Crossing Cuts]\label{def:conn-component-cuts}
	Given the set of $\eta$-near min cuts of a graph $G=(V,E)$, construct a graph where two cuts are connected by an edge if they cross. Partition this graph into maximal connected components. In the following, we will consider maximal connected components $\cC$ of crossing cuts and simply call them {\em connected components}.
	We say a connected component is a {\em singleton} if it has exactly one cut and a {\em non-singleton} otherwise. 
	
	For a connected component $\C$, let $\{a_i\}_{i \ge 0}$ be the coarsest partition of vertices $V$ such that for any $C\in \C$, either $a_i\subseteq C$ or $a_i\subseteq \overline{C}$. Each set $a_i$ is called an {\em atom} of $\C$ and we write  ${\cA}({\cal C})$ to denote the set of all atoms.
	
	Note for any atom $a_i \in \cA(\cC)$ which is an $\eta$-near min cut, $(a_i,\overline{a_i})$ is a singleton component, and is not crossed by any $\eta$-near min cut. Therefore $(a_i,\overline{a_i}) \not\in \cC$. 
	
	We can now represent any cut in $S \in \cC$ either by the set of vertices it contains or as a subset of $\cA(\cC)$. \textbf{In the following, we will often identify an atom with the set of vertices that it represents\footnote{For example, it will be convenient to write cuts as subsets of atoms. In this case the cut is the union of the vertices in those atoms.}.}
\end{definition}

 To study these systems, we will utilize the polygon representation of near minimum cuts of Bencz\'ur and Goemans \cite{Ben95,Ben97,BG08}. Their work implies that any connected component $\mathcal{C}$ of crossing $\eta$-near minimum cuts has a polygon representation with the following properties, so long as $\eta \le \frac{2}{5}$: 

\begin{figure}
\begin{center}
\begin{tikzpicture}[inner sep=1.7pt,scale=.7,pre/.style={<-,shorten <=2pt,>=stealth,thick}, post/.style={->,shorten >=1pt,>=stealth,thick}]
\tikzstyle{every node} = [draw, circle,color=black];
\begin{scope}[shift={(-5,0)}]
\foreach \i in {2,...,9}{
\path (\i*45:3) node  (a_\i) {\i};
}
\foreach \i in {0,...,7}{
\draw [color=blue,dashed,rotate around={22.5+45*\i:(22.5+\i*45:2.8)},line width=1.2] (22.5+\i*45:2.8) ellipse (.5 and 1.7);
}
\path (0,0) node (c) {1};
\foreach \a/\b in {2/3, 3/4, 4/5, 5/6, 6/7, 7/8, 8/9, 9/2}{
\path (a_\a) edge  (a_\b) edge [bend left=15] (a_\b) edge [bend right=15] (a_\b) edge (c);
}
\end{scope}
\begin{scope}[shift={(5,0)}]
\foreach \i in {2,...,9}{
\draw (22.5+\i*45:4) -- (22.5+45+\i*45:4);
\path (\i*45:3.4) node  (a_\i) {\i};
\draw [color=blue] (22.5+\i*45:4) -- (22.5+90+\i*45:4);
}
\path (0,0) node (c) {1};
\end{scope}
\end{tikzpicture}
\end{center}
\caption[A Family of near Minimum Cuts with an Inside Atom]{
Consider the graph on the left and suppose that every edge $e$ has fractional value $x_e =1/7$. This graph then has min cut value 2, with cuts of fractional value at most $2 + 1/7$ circled in blue. Note that this is a connected family $\C$ of near-min cuts, since every adjacent pair of blue cuts cross each other. The right image shows the polygon representation of $\C$. The blue lines in the right image are the representing diagonals. This representation has 8 outside atoms and $\{1\}$ is the only inside atom.} 
\label{fig:polygonrepresentation}
\end{figure}
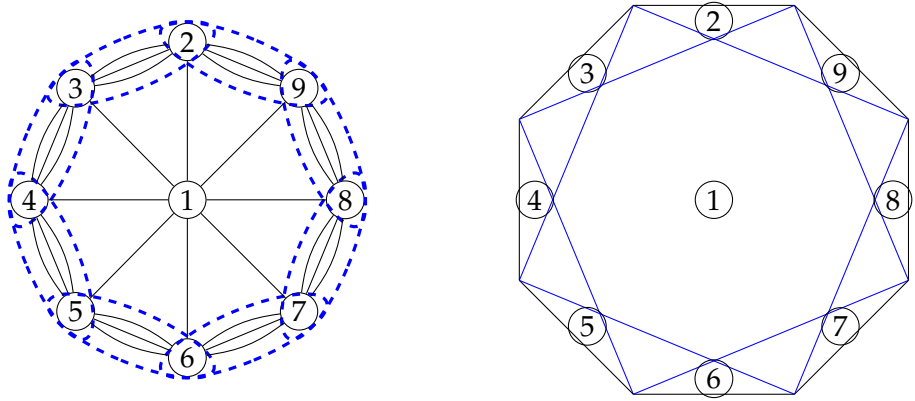

\begin{enumerate}
	\item A polygon representation is a convex regular polygon with a collection of \textit{representing diagonals}. All polygon edges and diagonals are drawn using straight lines in the plane. The diagonals partition the polygon into \textit{cells}. 
	\item Each atom of $\C$ is mapped to a cell of the polygon.  If one of these cells is bounded by some portion of the polygon boundary it is {\em non-empty} and we call its atom an \textit{outside atom}. We call the atoms of all other non-empty cells \textit{inside atoms}. Note that some cells may not contain any atom. WLOG label the outside atoms $a_0,\dots,a_{m-1}$ in counterclockwise order, and label the inside atoms arbitrarily. We also label points of the polygon $p_0,\dots,p_{m-1}$ such that outside atom $a_i$ is on the side $(p_i,p_{i+1})$ and $a_0$ is on the side $(p_{m-1},p_0)$. (In future sections we will refer to the special atom called the root, and if it is an outside atom WLOG we will label $a_0$ as the root.)
	\item No cell has more than one incident outer polygon edge.
	\item Each representing diagonal defines a cut such that each side of the cut is given by the union of the atoms on each side. Furthermore, the collection of cuts given by these diagonals is exactly $\mathcal{C}$. 
\end{enumerate}

The following fact follows immediately from the above discussion:
\begin{lemma}
\label{fact:atLeast2Outside}
Any cut $S\in \cC$ (represented by a diagonal of $P$) must have at least two outside atoms.
\end{lemma}

\begin{definition}[Outside atoms]
	For a polygon $P$ and a set $S$ of atoms of $P$, we write $O_P(S)$ to denote the set of outside atoms of $P$ in $S$; we drop the subscript when $P$ is clear from context. We also write $O(P)$ (or $O(\cA(\cC))$ where $\cC$ is the connected component of $P$) to denote the set of all outside atoms of $P$.
	
	Note that, given $S \in \cC$, since $S$ may be identified with a set of atoms, $O(S)$ is also well defined. 
\end{definition}

The following observation follows from  the fact that cuts correspond to straight diagonals in the plane and the polygon $P$ is regular:
\begin{property}[{\cite[Prop 19]{BG08}}] \label{obs:Ocross}
	If $S,S'\in {\cal C}$ cross then $O(S)$ and $O(S')$ cross, and $O(S \cup S') \not= O(P)$.
\end{property}

\begin{definition}[Left, Right Crossing]
Let $S,S'\in \cC$ such that $S'$ crosses $S$.
For such a pair, we say \textit{$S'$ crosses $S$ on the left} if the leftmost (clockwise-most) outside atom of $O(S' \cup S)$ is in $S'$. 
 Otherwise, we say that \textit{$S'$ crosses $S$ on the right}. Note that by \cref{obs:Ocross}, $O(S),O(S')$ cross. 
\end{definition}
\begin{definition}[Crossed on one, both sides]
\label{defn:crossedonetwo}
We say a cut $S$ is \textit{crossed on both sides} if it is crossed by a cut (in $\cC$) on the left and a cut (in $\cC$) on the right and we say $S$ is {\em crossed on one side} if it is crossed only on the left or only on the right. 
\end{definition}

 
\begin{definition}[Root node]\label{def:root}
	Recall $G_{/e_0}$ is the graph with $e_0$ contracted. Let $r \in V(G_{/e_0})$ be the result of contracting the nodes $\{u_0,v_0\}$.  We will call $r$ the \textbf{root node}.
\end{definition}

\begin{definition}[$\cN_\eta,\cN_{\eta,0},\cN_{\eta,1},\cN_{\eta,2},\cN_{\eta,\le 1}$]\label{def:cNeta}
	Given an LP solution $x$, let $\cN_\eta \subseteq 2^{V \smallsetminus \{r\}}$ be the set of all $\eta$-near min cuts of $x$ where we identify each cut with the side that does not contain the root node $r$.
	
	Let $\cN_{\eta,0} \subseteq \cN_\eta$ be the set of cuts that are not crossed. Let $\cN_{\eta,1} \subseteq \cN_\eta$ be the set of cuts that are crossed on one side in their respective polygons. Let $\cN_{\eta,2} \subseteq \cN_\eta$ be the set of cuts which are crossed on both sides in their respective polygons.  Finally let $\cN_{\eta,\le 1} = \cN_{\eta,0} \cup \cN_{\eta,1}$.
\end{definition}

\begin{mybox}{gray}{\hypertarget{preprocessing1}{Preprocessing Step 1: Compute the polygon representations}}
\begin{itemize}
	\item Find all $\eta$-near min cuts of the support graph $G_{/e_0}$, which can be done in deterministic polynomial time (for example see \cite{NNI94}).
	\item For each connected component of cuts $\cC$, compute its polygon representation $P$. By \cite{Ben95} this can be done in deterministic polynomial time.
	\item Given the collection of $\eta$-near min cuts and polygons, let $r$ be the root node and compute $\cN_\eta,\cN_{\eta, 0},\cN_{\eta,1},\cN_{\eta,2},$ and $\cN_{\eta,\le 1}$ (see \cref{def:cNeta}).
\end{itemize}
\end{mybox}

\subsection{Computation for cuts crossed on both sides}


\begin{definition}[Internal]\label{def:internal}
We say an edge is \textit{internal} to a polygon $P$ (of a connected component of cuts) if its endpoints fall into two different atoms of $P$, both of which are not the root atom of $P$. 
\end{definition}

Note by definition each edge is internal to at most one polygon $P$.

We iterate through each connected component of cuts $\cC$ in $\cN_\eta$ with polygon $P$ and do as follows. First, we define and compute: 

 \begin{definition}[$S_L$, $S_R$]\label{def:SLR}
 For each cut $S \in \cC$ which is crossed on both sides, let $S_L$ be the near minimum cut crossing $S$ on the left which minimizes $|O(S \cap S_L)|$. If there are multiple sets crossing $S$ on the left with the same minimum intersection, choose the smallest one to be $S_L$. Similarly, let $S_R$ be the near min cut crossing $S$ on the right which minimizes $|O(S \cap S_R)|$, and again choose the smallest set to break ties.
 \end{definition}
For each cut $S \in \cC$, we define:
\begin{equation}\label{eq:E-leftarrow-S}
\begin{split}
 	E^\leftarrow(S) &= E(S \cap S_L, S_L \smallsetminus S) \\
 	 E^\rightarrow(S) &= E(S \cap S_R, S_R \smallsetminus S) \\
 	 E^\circ(S) &= \delta(S) \smallsetminus (E^\leftarrow(S) \cup E^\rightarrow(S))
 	 \end{split}
 \end{equation}
In addition we define the left and right bad events for each polygon point $p$.
\begin{equation}\label{defn:badevents2}
	\begin{split}
		B^\rightarrow (p) &= \mathbb{1}\{|E^{\rightarrow} (L(p)) \cap T|\ne 1\text{ or }|E^\circ (L(p)) \cap T|\ne 0\}\\
B^\leftarrow (p) &= \mathbb{1}\{|E^{\leftarrow} (R(p)) \cap T|\ne 1\text{ or }|E^\circ (R(p)) \cap T|\ne 0\}.
	\end{split}
\end{equation}
If $L(p)$ does not exist, simply assume the left bad event never occurs, and similarly if $R(p)$ does not exist assume the right bad event never occurs.

Define $L(p)^{\cap R} := L(p) \cap L(p)_R$, and let $L^*(p) \in \cC$ be the cut crossing $L(p)^{\cap R}$ on the left that {\em maximizes}  $|O(L^*(p) \cap L(p)^{\cap R})|$ (and similarly $R^*(p)$ to maximize the intersection with $O(R(p)^{\cap L})$ on the right). If $L^*(p)$ does not exist, i.e. no cut crosses $L(p)^{\cap R}$ on the left, set $L^*(p) = \emptyset$, and similarly for $R^*(p)$. We let:
\begin{equation}\label{eq:bad-events-both-sides}
\begin{split}
	E(B^\rightarrow (p)) &:= E( L(p)^{\cap R} \smallsetminus L^*(p), L(p)_R \smallsetminus  L(p)^{\cap R})\\
	E(B^\leftarrow (p)) &:= E( R(p)^{\cap L} \smallsetminus R^*(p), R(p)_L \smallsetminus  R(p)^{\cap L})
	\end{split}
\end{equation}

\hypertarget{increase-bothsides-event}{
\begin{definition}[Increase event for cuts crossed on both sides]
	For each edge $e$ internal to polygon $P$, we define a random variable $\cI_e: \cT \to \{0,1\}$ which indicates if there exists a $p_i$ for which $e \in E(B^\rightarrow(p_i))$ and $B^\rightarrow(p_i)$ occurs or $e \in E(B^\leftarrow(p_i))$ and $B^\leftarrow(p_i)$ occurs.
\end{definition}}

In this way, $\cI_e$ has been defined for every edge internal to some polygon. For all edges $e$ which are not internal to any polygon, we simply let $\cI_e = 0$ for every tree. 

\begin{lemma}\label{lem:PIe}
We can compute
$$\P{\hyperlink{increase-bothsides-event}{\cI_e} \mid Set}$$
in polynomial time for any edge $e$. 
\end{lemma}
\begin{proof}
	If $e$ is not internal to any polygon, $\P{\cI_e \mid Set} = 0$ and we are done. Otherwise, it is internal to some polygon $P$. By Lemma 5.4 in \cite{KKO21b} (also see the proof of Theorem 5.2), there are at most two indices $i$ in this polygon $P$ for which $e \in E(B^\rightarrow(p_i))$ and at most two indices $i$ for which $e \in E(B^\leftarrow(p_i))$. Therefore, we are interested in at most four events $B^\rightarrow(p_i)$ or $B^\leftarrow(p_i)$.
	
	Using \cref{lem:master} it is straightforward to compute the probability that any collection of these (at most four) events occurs. For each event $i$ (say, some $B^\rightarrow(p_i)$) we use the two sets $E^\rightarrow(L(p_i))$ and $E^0(L(p))$ and set their $\sigma$ values to be 1 and 0 respectively and both of their $m$ values to be $|V|$, and return 1 minus the computed probability. 
	
	Therefore, we can compute the probability that at least one event occurs, which is sufficient to prove the lemma.
\end{proof}

Let $I_e: \cT_{partial} \to \R_{\ge 0}$ be the function from the above lemma which given $Set$ returns $\P{\hyperlink{increase-bothsides-event}{\cI_e} \mid Set}$.

\begin{mybox}{gray}{\hypertarget{preprocessing2}{Preprocessing Step 2: Compute polygon edge sets}}
For each polygon $P$ with connected component of cuts $\cC$:
\begin{itemize}
 	\item For each cut $S \in \cC$, compute $E^\leftarrow(S),E^\rightarrow(S),E^\circ(S)$ (see \cref{eq:E-leftarrow-S}).
 	\item For each polygon point $p$ in $P$, compute $E(B^\rightarrow(p))$ and $E(B^\leftarrow(p))$ (see \cref{eq:bad-events-both-sides}).
\end{itemize}
\end{mybox}

\begin{mybox}{gray}{\hypertarget{increaseboth}{\tt{Increase-Both-Sides}$(e,Set)$}}
Given an edge $e$ and $Set \in \cT_{partial}$ compute $\P{\hyperlink{increase-bothsides-event}{\cI_e} \mid Set}$ using \cref{lem:PIe}. 
\end{mybox}

\subsection{Preprocessing for cuts crossed on one side}

Now partition the cuts in $\cN_{\eta,1}$ into connected components. For each connected component of cuts $\cC$ and for each cut $C \in \cN_{\eta,0}$ that can be written as the union of two other cuts $a_1,a_2 \in \cN_{\eta, 0}$ which are not crossed, let $P$ be its (possibly degenerate)\footnote{In the case that $C \in \cN_{\eta,0}$, we simply let $P$ be the three atoms $a_0 = G \smallsetminus C$, $a_1,$ and $a_2$.} polygon $P$. Similar to above, each edge is internal (see \cref{def:internal}) to at most one such polygon $P$. By Lemma A.1 from \cite{KKO21b}, $P$ has no inside atoms. Label its outside atoms $a_0,\dots,a_{m-1}$, in counterclockwise order, where WLOG $a_0$ is the root atom.

We call $a_1$ the leftmost atom and $a_{m-1}$ the rightmost atom. Finally, for $1 \le i \le m$ let $\cE_i(P) = E(a_{i-1},a_{i \pmod{m}})$ be the edges between atom $a_{i-1}$ and $a_{i \pmod{m}}$ in $P$. 

Now we define the following:

\begin{definition}[$A,B,C$-Polygon Partition]\label{def:abcpolygonpartitioning}
The $A,B,C$-polygon partition of a polygon $P$ is a partition of edges of $\delta(a_0)$ into sets $A=\cE_1(P)$, $B=\cE_{m}(P)$, and $C=\delta(a_0)\smallsetminus A\smallsetminus B$. 
\end{definition}

\begin{definition}[Relevant Atoms and Relevant Cuts]\label{def:relevantcuts}
Define the family of relevant atoms of $\cC$ to be $$A=\{a_i:  1\leq i\leq m-1, x(\delta(a_i))\leq 2+\eta\},$$
and define the relevant cuts to be
$$\cC_{+}={\cal C}\cup A.$$
\end{definition}	

\begin{definition}[Left and Right  Hierarchies]
Let ${\cal L}$ (the {\em left  hierarchy}) be the set of all  cuts $A\in {\cal C}$ that are not crossed on the left. Similarly, we let ${\cal R}$ be the set of  cuts that are not crossed on the right. In this way ${\cal L},{\cal R}$ partition all cuts in ${\cal C}$. 
\end{definition}

Given this, define ${\cal C}_+^\cR = \cR \cup A$ and ${\cal C}_+^\cL = \cL \cup A$. 
\hypertarget{map}{\begin{definition}[$Map(\cE_{i}(P))$]
	We define a mapping from cuts in ${\cal C}_+^\cR$ to the edges $E(a_1,a_2),\dots,E(a_{m-2},a_{m-1})$. For any  $2\leq i\leq m-1$, we map 
\begin{equation}\label{eq:mapping-one-side}
	\argmax_{A\in \cC_+^\cR: \ell(A)=i} |A|\text{ and }\argmax_{A\in \cC_+^\cR: r(A)=i} |A|
\end{equation}
to $\cE_{i-1}(P)$, where $\ell(A)$ is the index of the leftmost atom of $A$ and $r(A)$ is the index of the rightmost atom of $A$. We then compute a similar mapping for $\cC_+^\cL$. For each edge group $\cE_{i}(P)$ we record the set of cuts mapped to it by these two processes as a multiset $Map(\cE_{i}(P))$ (since every atom is in both $\cC_+^R$ and $\cC_+^\cL$, some atoms may appear twice).
\end{definition}}

We now introduce the following notion:

\begin{definition}[Happy Cut]We say a leftmost cut $L\in \cC$  is {\em happy} if
$$ E(L, \overline{L\cup a_0})_T=1.$$
Similarly, the leftmost atom $a_1$ is {\em happy} if $E(a_1,\overline{a_0\cup a_1})_T=1$. Define  rightmost cuts in $\cC$ or the rightmost atom in $P$ to be happy in a similar manner.
\end{definition}

We now define an ``unhappy" event $\cU_C$ for each cut in $C \in \C_+$ such that 
$$\cU_C := \begin{cases}
 	\I{C \text{ is not happy}} & \text{If $C$ is a leftmost or rightmost cut} \\
 	\I{C \text{ is odd}} & \text{If $C$ is not a leftmost or rightmost cut}
 \end{cases}
$$

This allows us to define an increase random variable for each edge $e \in \cE_i(P)$ for $1 \le i \le m$ called $\cI'_e: \cT \to \{0,\frac{1}{2},1\}$. In particular for $e \in \cE_i(P)$ we let:
\hypertarget{increase-one-side}{$$\cI'_e := \min\{1, \sum_{C \in (Map(\cE_{i}(P)) \smallsetminus A)} \I{\cU_C} + \frac{1}{2}\sum_{C \in (Map(\cE_{i}(P)) \cap A)} \I{\cU_C}\}$$}
 where notice that an atom may contribute twice to the sum since $Map(\cE_{i}(P))$ may be a multiset.
 
 In this way, every edge which is internal to some polygon of $\cN_{\eta,\le 1}$ constructed in this section has an associated random variable $\cI'_e$. For every edge which is internal to no polygon constructed in this section, we say $\cI'_e$ never occurs.
 
\begin{lemma}  \label{lem:PI'e} 
We can compute
$$\P{{\cI'_e} \mid Set}$$
in polynomial time for any edge $e$ internal to a polygon $P$ of $\cN_{\eta,1}$. 
\end{lemma}
\begin{proof}
	If $e$ is not internal to any polygon of $\cN_{\eta,1}$ then $\P{\cI'_e \mid Set} = 0$ and we are done. Otherwise, it is internal to some polygon $P$ with root $a_0$. 
	Since by \cref{eq:mapping-one-side}, $|Map(\cE_i(P))| \le 4$, by linearity of expectation it is enough enough to compute $\P{\cU_C}$ for (at most four) cuts in $Map(\cE_{i}(P)$. We use \cref{lem:master}. Say $C$ is a leftmost cut  (it is similar if it is a rightmost cut). Then, compute $\P{\cU_C}=1-\P{C\text{ is happy}}$; so it is enough to compute $\P{C\text{ is happy} \mid Set}$. We use \cref{lem:master} with the set of edges $E(C,\overline{C \cup a_0})$ and with  corresponding $\sigma$ value of $1$ and $m$ value of $|V|$.
	If $C$ is not a leftmost or a rightmost cut we use \cref{lem:master} with the set $\delta(C)$, $\sigma$ value of $1$ and $m=2$.
\end{proof}

\begin{mybox}{gray}{\hypertarget{preprocessing3}{Preprocessing Step 3: Compute polygons of $\cN_{\eta,1}$ and the maps}}	
	Partition the cuts in $\cN_{\eta,1}$ into connected components. For each connected component of cuts $\cC$ compute its (possibly degenerate) polygon $P$. Now, for each polygon $P$ corresponding to a connected component $\C$ of cuts in $\cN_{\eta,1}$ with atoms $a_0,\dots,a_{m-1}$:
	\begin{itemize}
		\item Let $\cE_i(P)$ be the set of edges between $a_{i-1},a_{i\text{ mod }m}$
		\item	Let $\C_+$	be the set of relevant cuts as defined in \cref{def:relevantcuts}.
		\item For each $i\in 0,\dots,m-1$ construct the multiset \hyperlink{map}{$Map(\cE_i(P))$}  of cuts mapped to $\cE_i(P)$ in $\C_+$.
	\end{itemize}
%
\end{mybox}

\begin{mybox}{gray}{\hypertarget{increaseone}{\tt{Increase-One-Side}$(e,Set)$}}	
Given an edge $e$ and $Set \in \cT_{partial}$ compute $\P{\hyperlink{increase-oneside-event}{\cI'_e} \mid Set}$ using \cref{lem:PI'e}. 
\end{mybox}

\subsection{Computation of $\E{c(s^*) \mid Set}$}







Following Theorem 6.1 of \cite{KKO21b} we define $s^*_e: \cT \to \R_{\ge 0}$:
\begin{equation}\label{eq:s*e}s^*_e = (1-\gamma) \frac{2+\eta}{1-\eps_\eta} \decrease x_e (\cI_e + \cI'_e) + \gamma 2\decrease x_e \cI_e,	
\end{equation}
where $\gamma = \frac{15}{32}\eps_P$. By the above two lemmas, we can compute $\E{c(s^*) \mid Set}$ in polynomial time. Thus the fact that the following function can be computed efficiently is the main result of this section:

\begin{mybox}{yellow}{\hypertarget{csstar}{$\mathbb{E}_{c(s^*)}(e,Set)$}}
	Given an edge $e$ and $Set\in \cT_{partial}$, call functions 
	\hyperlink{increaseone}{{\tt Increase-One-Side}$(e,Set)$} and \hyperlink{increaseboth}{{\tt Increase-Both-Sides}$(e,Set)$} to compute $\P{I'_e \mid Set}$ and $\P{I_e \mid Set}$ respectively. Then use \eqref{eq:s*e} to compute and return $\E{c(s^*_e) \mid Set}$.
\end{mybox}
This concludes the proof of (1) of \cref{lem:compute-s}.

%% file: overview-laminar.tex
\section{Computation for $c(s)$}\label{sec:preprocessing-s}

Here we will compute some parameters which are fixed throughout the course of the algorithm. We also classify edges based on the probability of some events. In all computations we use the (unconditional) measure $\mu_\lambda$. 

We begin by setting the constants as in \cref{table:constants}. 
\begin{table}\centering
\begin{tabular}{|c|c|c|c|}
	\hline Name & Value & Explanation\\
	\hline $\eps_{1/2}$ & 0.0002 & Half edge threshold\\
	\hline $\eps_{1/1}$ & $\frac{\eps_{1/2}}{12}$ & $A,B,C$ partitioning threshold, \cref{def:abcdegpartitioning} \\
	\hline $p$ & $0.005\eps_{1/2}^2$ & Min prob. of happiness for a (2-*) good edge \\
	\hline $\eps_M$ & $0.00025$ & Marginal errors due to max flow \\
	\hline $\tau$ & $0.571\beta$ & Top edge decrease \\ 
	\hline $\eps_P$ & $750\eta$ & Expected decrease constant\\
	\hline $\alpha$ & $2\eps_\eta$ & Parameter of the matching \\ 
	\hline $\eps_B$ & $21\eps_{1/2}$ & Parameter of the matching \\
	\hline $\eps_F$ & $1/10$ & Parameter of the matching \\
	\hline $\eps_\eta$ & $7\eta$ & \cref{def:hierarchy} \\
	\hline $\eta$ & $4.16 \cdot 10^{-19}$ & Near min cut constant \\
	\hline $\decrease$ & $\frac{\eta}{4+2\eta}$ & Slack shift constant\\
	\hline 
\end{tabular}
\caption{A table of all constants used in the paper.}\label{table:constants}
\end{table}

\subsection{Hierarchy definition and computation}\label{subsec:hierarchy}

Here we recall notation from \cite{KKO21b}. The following is key to defining the slack vector. 

\begin{definition}[Hierarchy, \cite{KKO21b}]\label{def:hierarchy}
\hypertarget{tar:hierarchy}{For an LP solution $x^0$ with support $E_0=E\cup \{e_0\}$ where $x$ is $x^0$ restricted to $E$, a hierarchy ${\cal H}\subseteq \cN_{\eps_\eta}$ is a {\em laminar} family  with root $V\smallsetminus \{u_0,v_0\}$, where every cut $S\in \cH$ is called either a ``near-cycle" cut or a degree cut. In the special case that $S$ has exactly two children we call it a triangle cut. Furthermore, every cut $S$ is the union of its children. 
For any (non-root) cut $S\in \cH$, define the parent of $S$, $\p(S)$, to be the smallest cut $S'\in\cH$ such that $S\subsetneq S'$.}

\hypertarget{tar:AS}{For a cut $S\in \cH$, 
let $\cA(S):=\{a\in \cH: \p(a)=S\}$; we will call these the atoms of $S$. If $S$ is called a ``near-cycle" cut, then we can order cuts in $\cA(S)$, $a_1,\dots,a_{m-1}$ such that 
\begin{itemize}
\item $x(E(\overline{S},a_1)), x(E(a_{m-1},\overline{S})) \ge 1-\eps_\eta$.
\item For any $1\leq i<m-1$, $x(E(a_i,a_{i+1}))\geq 1-\eps_\eta$.
\item $\cup_{i=2}^{m-2} E(a_i,\overline{S}) \le \eps_\eta$.
\end{itemize}}

We abuse notation and for an  edge $e=(u,v)$ that is not a neighbor of $u_0,v_0$, we write $\p(e)$ to denote the smallest\footnote{in the sense of the number of vertices that it contains} cut $S'\in \cH$ such that $u,v\in S'$.
\end{definition}

\begin{definition}[$A,B,C$ near-cycle partition, left-happy, right-happy, and happy]\label{def:ABC-leftrighthappy}
	Let $\cH$ be a hierarchy and let $S \in \cH$ be a near-cycle cut with cuts in $\cA(S)$ ordered $a_1,\dots,a_{m-1}$. Then let $A=x(E(\overline{S},a_1))$, $B=x(E(a_{m-1},\overline{S}))$, and $C=\cup_{i=2}^{m-2} E(a_i,\overline{S})$. We call the sets $A,B,C$ the \textbf{near-cycle partition} of $\delta(S)$. 
	
	We say $S$ is \textbf{left-happy} when $A_T$ is odd and  $C_T=0$, \textbf{right happy} when $B_T$ is odd and $C_T=0$, and \textbf{happy} when $A_T,B_T$ are odd and $C_T=0$.
	
	By \cref{def:hierarchy}, we have $x(A),x(B) \ge 1-\eps_\eta$ and $x(C) \le \eps_\eta$.
\end{definition}

%

Now we will define a hierarchy $\cH$ as the cuts in $\cN_{\eta}$ which are not crossed, plus some extra cuts in $\cH_{\eps_\eta}$. 

\begin{mybox}{gray}{\hypertarget{preprocessing4}{Preprocessing Step 4: Constructing the hierarchy}}
Let $\cN_{\eta,\le 1} \subseteq 2^{V \smallsetminus \{u_0,v_0\}}$ be the set of cuts crossed on at most one side: this was computed in \cref{sec:preprocessing-sstar}. Now we construct $\cH$ as follows. For every connected component ${\cal C}$ of $\cN_{\eta,\le 1}$, if $|{\cal C}|=1$ then add the unique cut in ${\cal C}$ to the hierarchy. Otherwise, ${\cal C}$ corresponds to a connected component of cuts crossed on one side $u$ with atoms $a_0,\dots,a_{m-1}$ (for some $m>3$). Add $a_1,\dots,a_{m-1}$\footnote{Notice that an atom may already correspond to a connected component, in such a case we do not add it in this step.} and $\cup_{i=1}^{m-1} a_i$ to $\cH$.  
Note that since $x(\delta(\{u_0,v_0\}))=2$, the root of the hierarchy is always $V\smallsetminus \{u_0,v_0\}$. \\

Now, partition the cuts in $\cH$ into degree cuts and near-cycle cuts. For a cut $S \in \cH$, if there is a connected component of at least two cuts  with union equal to $S$, then call $S$  a \textbf{near-cycle cut} and compute its near-cycle $A,B,C$ partitioning as defined in \cref{def:ABC-leftrighthappy}. 
If $S$ is a cut with exactly two children $X,Y$ in the hierarchy, then also call $S$ a near-cycle cut\footnote{Think about such a set as a {\em degenerate} polygon with atoms $a_1:= X,a_2:= Y,a_0 :=\overline{X\cup Y}$.}, with  $A,B,C$ partitioning $A=E(X,\overline{X}\smallsetminus Y)$, $B=E(Y,\overline{Y}\smallsetminus X)$ and $C=\emptyset$.
Otherwise, call $S$ a \textbf{degree cut}. \\

Finally, compute the $A,B,C$ degree partitioning for all $S \in \cH$ as described below in \cref{def:abcdegpartitioning}.
\end{mybox}

\begin{remark}
Since $|\cN_{\eta,\le 1}|$ has polynomial size in $n$ this can be done in polynomial time.

Also note that since every vertex has degree 2, they all appear in the hierarchy as singletons. Therefore, every set in the hierarchy is the union of its children.	
\end{remark}

\subsection{Edge bundles, $A,B,C$ degree partition, and edge classification}\label{subsec:bundles}


\begin{definition}[Edge Bundles, Top Edges, and Bottom Edges]\label{def:edgebundletopcut}
 For every degree cut $S$ and every pair of atoms $u,v\in\hyperlink{tar:AS}{\cA(S)}$,  we define a \textbf{top edge bundle} $\bbf=(u,v)$ such that 
	$$\bbf= \{e=(u',v') \in E : \p(e) = S, u'\in u, v'\in v\}.$$
	 Note that in the above definition, $u',v'$ are actual vertices of $G$.
	 
For every polygon cut $S$, we define the \textbf{bottom edge bundle} $\bbf=\{e: \p(e)=S\}$. 

Note in this way every edge $e$ is in a unique edge bundle $\bbe$. We say $e$ is a bottom edge if its edge bundle is a bottom edge bundle and otherwise $e$ is a top edge.
\end{definition}
We will always use bold letters to distinguish top edge bundles from actual LP edges. Also, we abuse notation and write $x_\bbe:=\sum_{f\in\bbe}x_f$ to denote the total fractional value of all edges in this bundle.


 For any $u\in\cH$ with $\p(u)=S$ we write
\begin{align} 
\delta^\uparrow(u)&:=\delta(u)\cap\delta(S),\nonumber\\
\delta^\rightarrow(u)&:=\delta(u)\smallsetminus\delta(S).\label{def:arrows} \\
\hypertarget{tar:ErightarrowS}{E^\rightarrow (S)&:= \bigcup_{v \in \cH: \p(v)=S} \delta^\rightarrow(v)}.\nonumber
\end{align}
Also, for a set of edges $A\subseteq \delta(u)$ we write $A^\rightarrow, A^\uparrow$ to denote $A\cap\delta^\rightarrow(u), A\cap\delta^\uparrow(u)$ respectively (when $u$ is clear in context).
Note that $E^\rightarrow(S)\subseteq E(S)$ includes only edges between atoms of $S$ and not all edges between vertices in $S$.  




Now we compute the so-called $A,B,C$ degree partitioning of each cut $S \in \cH$ for which $p(S)$ is a degree cut. It can easily be implemented in polynomial time.

\begin{definition}[$A,B,C$-Degree Partitioning]\label{def:abcdegpartitioning}
\hypertarget{tar:degreepartition}{For $u\in\cH$ and $\eps_{1/1}$ as in \cref{table:constants}, we define a partitioning of edges in $\delta(u)$: Let $a,b\subsetneq u$ be {\em minimal} cuts in the hierarchy, i.e., $a,b\in\cH$, such that $a\neq b$  and $x(\delta(a)\cap\delta(u)),x(\delta(b)\cap\delta(u))\geq 1-\eps_{1/1}$. Note that since the hierarchy is laminar, $a,b$ cannot cross. Let $A=\delta(a)\cap\delta(u),B=\delta(b)\cap\delta(u),C=\delta(u)\smallsetminus A\smallsetminus B$. }

If there is no cut $a\subsetneq u$ (in the hierarchy) such that $x(\delta(a)\cap\delta(u))\geq 1-\eps_{1/1}$, we just let $A,B$ partition $\delta(u)$ such that $x(A),x(B) \in [1-\eps_{1,1},1+\eps_\eta]$, and set $C=\emptyset$. Note that this exists WLOG because we may split any edge into an arbitrary number of parallel copies.

If there is just one minimal cut $a\subsetneq u$ (in the hierarchy) with $x(\delta(a)\cap\delta(u))\geq 1-\eps_{1/1}$, i.e., $b$ does not exist in the above definition, then we define $A=\delta(a)\cap\delta(u)$. Let $a'\in\cH$ be the unique child of $u$ such that $a\subseteq a'$, i.e., $a$ is equal to $a'$ or a descendant of $a'$. Then we define $B$ to be an arbitrary subset of $\delta(u) \smallsetminus \delta(a')$ such that $x(B) \in  [1-\eps_{1,1},1+\eps_\eta]$. Finally let $C = \delta(u) \smallsetminus (A \cup B)$. Note $C \supseteq \delta(a')\cap\delta(u) \smallsetminus \delta(a)$.
\end{definition}

Note we may have to divide a single edge $e$ between the sets $A,B,C$ to ensure such partitions exist. 



\begin{table}\centering
\begin{tabular}{|c|c|c|c|}
	\hline Variable & Name & Event\\
	 \hline  $H_\bbe$ & 2-2 happy & $\P{u,v \text{ trees}, \delta(u)_T=\delta(v)_T=2}$\\
	\hline  $H_{\bbe,u}$ & 2-1-1 happy w.r.t $u$ & $\P{u,v \text{ trees}, A_T=B_T=1, C_T=0, \delta(v)_T=2}$\\
	\hline $H_{\{\bbe,\bbf\}}$ & 2-2-2 happy w.r.t $u$ with $\bbf=(u,w)$ &  $\P{u,v,w \text{ trees}, \delta(u)_T=\delta(v)_T=\delta(w)_T=2}$\\
	\hline 
\end{tabular}
\caption{For a top edge bundle $\bbe=(u,v)$ where $A,B,C$ is the degree partitioning of $u$, we define the following ``happy" events.}\label{table:events}
\end{table}

Let $p$ be as in \cref{table:constants}. For a top edge bundles $\bbe=(u,v)$, we say $\bbe$ is \textbf{2-2 happy}, or $H_\bbe$ occurs, if $u,v$ are trees and $\delta(u)_T = \delta(v)_T = 2$. 
Recall that $u,v \in \cH$ are sets of vertices. 
\begin{mybox}{gray}{\hypertarget{p22}{{\tt $p_{\text{2-2}}(\bbe,Set)$}}}
	To compute \hyperlink{22happy}{$\P{\bbe \text{ 2-2 happy}}$}, use \cref{lem:master} with 
$$E_1 = E(u), \quad E_2 = E(v), \quad E_3 = \delta(u), \quad E_4 = \delta(v),$$
$$\vec{\sigma}=(|u|-1,|v|-1,2,2), \quad\quad \vec{m}=(|V|,|V|,|V|,|V|),$$
\end{mybox}
\begin{definition}[Good and bad edges]
	A top edge $e$ in edge bundle $\bbf$ is \textbf{good} (sometimes just ``good") if $p_{2-2}(\bbf,\emptyset) \ge p$ and \textbf{bad} otherwise. We say every bottom edge is good, and edges in $\delta(\{u_0,v_0\})$ are bad (because they do not have both of their endpoints in the hierarchy).
\end{definition}


\begin{itemize}
\item  \textbf{2-1-1 happy w.r.t. $u$:} Let $A,B,C$ be the $A,B,C$ degree partition of $u$ computed in the previous section. \hypertarget{211happy}{We say $\bbe$ is 2-1-1 happy w.r.t. $u$, or $H_{\bbe,u}$ occurs, if $u,v$ are trees, $A_T = B_T = 1, C_T=0,$ and $\delta(v)_T = 2$.}

\begin{mybox}{gray}{\hypertarget{p211}{{\tt $p_{\text{2-1-1}}(\bbe,u,Set)$}}}
	Let $A,B,C$ be the degree partition of $u$. To compute $\P{\hyperlink{211happy}{\bbe \text{ 2-1-1 happy w.r.t } u}}$,
	use \cref{lem:master} with $E_1 = E(u), E_2 = E(v)$, $E_3 = A, E_4 = B, E_5 = C, E_6 = \delta(v)$, $\vec{\sigma}=(|u|-1,|v|-1,1,1,0,2)$, and $\vec{m}=(|V|,|V|,|V|,|V|,|V|,|V|).$ 
\end{mybox}

\item \hypertarget{222happy}{\textbf{2-2-2 happy (w.r.t. common endpoint $u$, with partner $f$)}: We say the edge bundles $\bbe=(u,v)$ and $\bbf =(u,w)$ (where $\p(u) = \p(v) = \p(w)$) are 2-2-2 happy w.r.t. $u$, or $H_{\{\bbe,\bbf\}}$ occurs, if $u,v,w$ are trees and $\delta(u)_T = \delta(v)_T = \delta(w)_T = 2$.} 

\begin{mybox}{gray}{\hypertarget{p222}{{\tt $p_{\text{2-2-2}}(\bbe,\bbf,Set)$}}}
	Assume $\bbe,\bbf$ have a common endpoint $u$ and $v,w$ are the other endpoints of $\bbe,\bbf$. 

	To compute $\P{\hyperlink{222happy}{\bbe \text{ 2-2-2 happy w.r.t $u$ with $\bbf$}}}$,
	use \cref{lem:master} with $E_1 = E(u), E_2 = E(v), E_3 = E(w)$, $E_4 = \delta(u), E_5 = \delta(v), E_6 = \delta(w)$, $\vec{\sigma}=(|u|-1,|v|-1,|w|-1,2,2,2)$, and $\vec{m}=(|V|,|V|,|V|,|V|,|V|,|V|).$
\end{mybox}\end{itemize}


For each edge bundle $\bbe=(u,v)$, we define its type with respect to each endpoint as follows:
\begin{mybox}{gray}{\hypertarget{type}{{\tt type}$(u)$} \hspace*{60mm} returns {\tt type}$(\bbe,u)$ for all edge bundles $\bbe \in \delta(u)$}
	For every edge bundle $\bbe \in \delta(u)$:

	\hspace*{3mm} If $\hyperlink{p211}{p_{\text{2-1-1}}(\bbe,u,\emptyset)} \ge p$, then {\tt type}$(\bbe,u) = \text{2-1-1}$. \\
	
	Let $A,B,C$ be the degree partitioning of $\delta(u)$. Let $F \subseteq \delta(u)$ be the set of edge bundles adjacent to $u$ with {\tt type}$(\bbe,u) = \text{2-1-1}$. Let $J \subseteq \delta(u)$ be the set of edges adjacent to $u$ such that $\hyperlink{p22}{p_{\text{2-2}}(\bbe,\emptyset)} < p$. Now, if $x(F) \le \frac{1}{2}-\eps_{1/2}-\eps_\eta$ and $x(J) \le \frac{1}{2}-\eps_{1/2}$, then, by Theorem 5.28 of \cite{KKO21a}, there exists two edges $\bbe=(u,v),\bbf=(u,w)$ such that: (i) $\hyperlink{p222}{p_{\text{2-2-2}}(\bbe,\bbf,\emptyset)} \ge p$, and (ii) $x(\bbe \cap B) \le \eps_{1/2}$, $x(\bbf \cap A) \le \eps_{1/2}$. Let {\tt type}$(\bbe,u) = $ {\tt type}$(\bbf,u) =\text{2-2-2}$. \\

	For every edge bundle $\bbe \in \delta(u)$ such that {\tt type}$(\bbe,u)$ is not set: \\
	\hspace*{3mm} If $p_{\text{2-2}}(\bbe,\emptyset) \ge p$, then {\tt type}$(\bbe,u)=\text{2-2}$. Otherwise {\tt type}$(\bbe,u)=\text{bad}$. 
\end{mybox}

Finally, for each edge bundle $\bbe=(u,v)$, we define the following $\cR_{\bbe,u}$ event with respect to each endpoint $u$:
\begin{itemize}
\item If {\tt type}$(\bbe,u)=\text{2-1-1}$, we define an independent Bernoulli $B_{\bbe,u}$ with success probability $p/p_{2-1-1}(\bbe,u,\emptyset)$ and we define $$\cR_{\bbe,u} := \I{H_{\bbe,u} = B_{\bbe,u} = 1}$$

We emphasize that this reduction indicator is purely a function of a tree $T$ and an independent Bernoulli. The same will apply to all future reduction indicators. 
\item If {\tt type}$(\bbe,u)=\text{2-2-2}$, there exists an edge bundle $\bbf$ such that {\tt type}$(\bbf,u)=\text{2-2-2}$. In this case, define an independent Bernoulli $B_{\{\bbe,\bbf\}}$ with success probability $p/p_{2-2-2}(\bbe,\bbf,\emptyset)$. Define
$$\cR_{\bbe,u} := \cR_{\bbf,u} := \cR_{\{\bbe,\bbf\}} := \I{H_{\{\bbe,\bbf\}} = B_{\{\bbe,\bbf\}} = 1}$$

Note we use brackets to emphasize that $\cR_{\{\bbe,\bbf\}}$ is the same event as $\cR_{\{\bbf,\bbe\}}$.

\item If {\tt type}$(\bbe,u)=\text{2-2}$. We define an independent Bernoulli $B_{\bbe}$ with success probability $p/p_{2-2}(\bbe,\emptyset)$ and we define $$\cR_{\bbe,u} := \I{H_\bbe = B_\bbe = 1}.$$
\item Otherwise {\tt type}$(\bbe,u)=\text{bad}$. Define $\cR_{\bbe,u} = 0$.
\end{itemize}

%
%
%
The following allows us to compute the expected value of $\cR_{\bbe,u}$ for all edge bundles and each of their endpoints conditioned on $Set$. Note that $\P{\cR_{\bbe,u}} = p$ for all good edges. However, $\P{\cR_{\bbe,u} \mid Set}$ can be any number between 0 and the success probability of its Bernoulli (defined above). 
\begin{mybox}{gray}{\hypertarget{ReSet}{$\mathbb{E}_{\cR}(\bbe,u,Set)$}}
	Call \hyperlink{type}{{\tt type}$(u)$} to determine {\tt type}$(\bbe,u)$. If {\tt type}$(\bbe,u)=\text{2-1-1}$, return $$\hyperlink{p211}{p_{\text{2-1-1}}(\bbe,u,Set)} \cdot \left(\frac{p}{p_{\text{2-1-1}}(\bbe,u,\emptyset)}\right)$$
	Otherwise if {\tt type}$(\bbe,u)$ = {\tt type}$(\bbf,u)=\text{2-2-2}$ for some edge bundle $\bbf$, return
	$$\hyperlink{p222}{p_{\text{2-2-2}}(\bbe,\bbf,Set)} \cdot \left(\frac{p}{p_{\text{2-2-2}}(\bbe,\bbf,\emptyset)}\right)$$
	Otherwise, if {\tt type}$(\bbe,u)=\text{2-2}$, return
	$$\hyperlink{p22}{p_{\text{2-2}}(\bbe,Set)} \cdot \left(\frac{p}{p_{\text{2-2}}(\bbe,\emptyset)}\right)$$
	Otherwise, {\tt type}$(\bbe,u)=\text{bad}$. Return 0.
\end{mybox}


%

\subsection{Max Flow}\label{subsec:maxflow}

For each near-cycle cut $S \in \cH$ with polygon partition $A,B,C$, we compute parameters $\alpha_{e,f}$ for all edges $e \in A,f \in B$ as well as $p_S$, which we define next. Let $H_S$ be the event: 
$$H_S:= A_T=B_T=1,C_T=0, S \text{ is a tree}.$$

\begin{mybox}{gray}{\hypertarget{preprocessing5}{Preprocessing Step 5: Max Flow}}
For every near-cycle cut $S \in \cH$ with polygon partition $A,B,C$, do the following. \\

	Construct and solve an instance of the max-flow, min-cut problem. Consider the following graph with vertex set $\{s,A,B,t\}$.  
For any edge $e\in A, f\in B$ connect $e$ to $f$ with a directed edge of capacity $y_{e,f}=\P{e,f\in T \mid H_S} = \frac{\P{H_S \land e,f\in T}}{\P{H_S}}$. To compute the numerator (the denominator is similar), apply \cref{lem:master} to 
\begin{equation}\label{eq:R_S}
\begin{aligned}
	&E_1 = \{e\}, \quad E_2 = \{f\}, \quad E_3 = \delta(S) \smallsetminus \{e,f\}, \quad E_4 = E(S) \\
	&\quad \quad \quad \vec{\sigma} = (1,1,0,|S|-1), \quad \vec{m} = (2,2,|V|,|V|)
\end{aligned}
\end{equation}

For any $e \in E$, let $x_e := \P{e \in T \mid C_T = 0, S \text{ is a tree}}$. Connect $s$ to $e\in A$ with an arc of capacity $q x_e$ and similarly connect $f\in B$ to $t$ with arc of capacity $q x_f$, where $q =\frac{0.1\zeta^2/6^2}{\P{A_T=B_T=1 \mid C_T=0,S \text{ is a tree}}}$ and $\zeta=1/4000$, computed using \cref{lem:master}. Then compute the maximum flow of this graph. Let $\bf{z}$ be the maximum flow, where $z_{e,f}$ is the flow on the edge from $e$ to $f$. \\

Now return 
\begin{equation}\label{eq:alpha,pS}
	\hypertarget{alphaef}{\alpha_{e,f} := \frac{z_{e,f}}{y_{e,f}}, \quad\quad\quad\quad p_S := \sum_{e \in A, f \in B} \P{H_S}z_{e,f}}
\end{equation}
\end{mybox}

Note that it follows by Proposition 5.6 from \cite{KKO21a} that $p_S \ge p$. Define an independent Bernoulli $B_S$ with success probability $p/p_S$ as well as independent Bernoullis $B_{e,f}$ for all $e \in A, f \in B$ with success probability $\alpha_{e,f}$. For a tree $T$, we define the event:
\begin{equation}\label{eq:defRS-maxflow}
\cR_S := \bigcup_{e \in A, f \in B} \I{A \cap T = \{e\}, B \cap T = \{f\}, H_S= B_S= B_{e,f}=1}
\end{equation}

\begin{mybox}{gray}{\hypertarget{RpolySet}{$\mathbb{E}_{\cR}(S,Set)$}}
To compute $\E{\cR_S \mid Set}$, note by definition:
 \begin{align}
 	\E{\cR_S \mid Set} &= \sum_{f \in A, g \in B} \P{A \cap T = \{f\}, B \cap T = \{g\}, H_S = B_S = B_{e,f} = 1 \mid Set}\nonumber \\
 	&= (p/\hyperlink{alphaef}{p_S})\hyperlink{alphaef}{\alpha_{e,f}} \sum_{f \in A, g \in B} \P{A \cap T = \{f\}, B \cap T = \{g\}, H_S \mid Set},\label{eq:rpolygon}
 \end{align}

 
 
Compute the inner probability using \cref{lem:master} similarly to \eqref{eq:R_S}. Return the result.
\end{mybox}




\subsection{Matching}\label{subsec:matching}

Next we compute a matching from the good edges in $E^\rightarrow(S)$ to the edges in $\delta(S)$ for every degree cut $S \in \cH$. The output of this procedure will be values $m_{\bbe,u}$ which indicate that the good edge bundle $\bbe={\bf(u,v)}$ (where $u,v\in\hyperlink{tar:AS}{\cA(S)}$) is matched to a fraction $m_{\bbe,u}$ of edges in $\delta^\uparrow(u)$ and a fraction $m_{\bbe,v}$ of $\delta^\uparrow(v)$. In the following, $\eps_{F},\eps_B$, and $\alpha$ are set in \cref{table:constants}. 

\begin{mybox}{gray}{\hypertarget{preprocessing6}{Preprocessing Step 6: Matching}}
For every $S \in \cH$ which is a degree cut, do the following. For every $u \in \hyperlink{tar:AS}{\cA(S)}$, set:
$$\hypertarget{tar:Fu}{F_u=1-\eps_B\I {\eps_F \le x(\delta^\uparrow(u)) \le 1-\eps_F},} \quad
\hypertarget{tar:Zu}{Z_u:=\left(1 + \I{|\cA(S)|\geq 4, x(\delta^\uparrow(u))\leq \eps_F}\right).}$$

Set up and solve a polynomial size max-flow min-cut problem. Construct a graph with vertex set $\{s,X,Y,t\}$ with source $s$ and sink $t$. We identify $X$ with the set of good edge bundles in $E^\rightarrow(S)$ and $Y$ with the set of atoms in $\cA(S)$. \\

For every (good) edge bundle $\bbe \in X$, add an arc from $s$ to $\bbe$ of capacity $c(s,\bbe):= (1+\alpha) x_\bbe$.  For every $u\in \cA(S)$, 
	add an arc $(u,t)$ with capacity $c(u,t) = x(\delta^\uparrow(u))F_u Z_u.$
Finally, connect $\bbe=(u,v)\in X$ to each of $u,v\in Y$ with a directed edge of infinite capacity, i.e., $c(\bbe,u)=c(\bbe,v)=\infty$. \\

Let $f$ be the max flow and return 
\begin{equation}\label{eq:matching}
m_{\bbe,u} := \frac{f_{\bbe,u}}{F_u}.
\end{equation}
\end{mybox}

\subsection{Reductions}

In the following we compute the probability of events $\cR$, corresponding to the probability of ``decrease events" for every edge bundle. These then are used to compute values $r_e$ for every edge, corresponding to the actual decrease amounts. For a set $F \subseteq E$, we let $r(F) = \sum_{e \in F} r_e$. 

If $e$ is a top edge (see \cref{def:edgebundletopcut}), then $e \in \bbf$ for some top edge bundle $\bbf=(u,v)$. Define
$$r_e = \frac{1}{2}\tau x_e (\I{\cR_{\bbf,u}} + \I{\cR_{\bbf,v}})$$
If $\bbe$ is a bottom edge with near-cycle parent $S$, then define
$$r_e = \beta x_e \I{\cR_S}$$
Where $\tau,\beta$ are given in \cref{table:constants}.

\begin{mybox}{gray}{\hypertarget{littlereset}{$\mathbb{E}_{r}(e,Set)$}}
	If $e$ is a top edge in top edge bundle $\bbf=(u,v)$ return 
	$$\E{r_e \mid Set} = \frac{1}{2}\tau x_e (\hyperlink{ReSet}{\mathbb{E}_{\cR}(\bbf,u,Set)} + \hyperlink{ReSet}{\mathbb{E}_{\cR}(\bbf,v,Set)})$$
	Otherwise, $e$ is a bottom edge with near-cycle parent $S$. Return 
 $$\E{r_e \mid Set} = \beta x_e \hyperlink{RpolySet}{\mathbb{E}_{\cR}(S,Set)}.$$ 
	
\end{mybox}

\subsection{Increases}

We now recall the definition of increase vectors in \cite{KKO21a} (over all edges) with the purpose of guaranteeing that every odd cut in $\cH$ is satisfied and then show how to compute its expectation. In the following subsection, the slack vectors are defined as the sum of the decrease vector and (a scaled version of) the increase vector. 


\subsubsection{Increases for bottom edges}

Here we define the increase needed for bottom edges in each near-cycle cut $S$ with near-cycle partition $A,B,C$. We let $I_S = I_S^\uparrow + I_S^\rightarrow$. We define:
\begin{align}\label{def:Iup}
	I_S^\uparrow = (1+\eps_\eta)(r(A^\uparrow)\I{S \text{ not left happy}} + r(B^\uparrow)\cdot \I{S \text{ not right happy}} + r(C))
\end{align}





\begin{mybox}{gray}{\hypertarget{IupSet}{$\mathbb{E}_{I^\uparrow}(S,Set)$}}
	Assume $S$ is a near-cycle cut with near-cycle partition $A,B,C$. To compute the expected value of the first term in \eqref{def:Iup}, note by linearity of expectation it suffices to compute the following for each $e \in A^\uparrow$. 
\begin{itemize}
\item If $e$ is a top edge in top edge bundle $\bbf=(u,v)$, compute:
$$\E{r_e \cdot \I{S \text{ not left happy}} \mid Set} = \frac{1}{2} x_e \tau \E{(\cR_{\bbf,u} + \cR_{\bbf,v}) \cdot \I{S \text{ not left happy}} \mid Set},$$
The expectation is equivalent to $\hyperlink{ReSet}{\mathbb{E}_{\cR}(\bbf,u,Set)} - \E{\cR_{\bbf,u} \land \I{S \text{ left happy}} \mid Set}$ (plus the analogous quantity for $v$). To compute the second term, first recall the definition of left happy (see \cref{def:ABC-leftrighthappy}): $A_T$ is odd and $C_T = 0$. Now apply \cref{lem:master} using the necessary sets $E_i$ and vectors $\vec{m},\vec{\sigma}$ for $\cR_{\bbf,u}$ (as given in $\hyperlink{ReSet}{\mathbb{E}_{\cR}(\bbf,u,Set)}$) and add two additional sets $E_A = A$ and $E_C = C$ and coordinates $\sigma_A = 1, \sigma_C = 0, m_A = 2, m_C = |V|$. (Note if $\bbf$ is bad then the whole expectation is 0 and there is nothing to compute.)
\item If instead $e$ is a bottom edge in near-cycle cut $\hat{S}$, similarly compute:
$$\E{r_e \cdot \I{S \text{ not left happy}} \mid Set} =  x_e \decrease \E{(\cR_{\hat{S}} \cdot \I{S \text{ not left happy}} \mid Set}.$$
Here to apply \cref{lem:master}, sum over the events for $\cR_{\hat{S}}$ as in \eqref{eq:rpolygon} (used by \hyperlink{RpolySet}{$\mathbb{E}_{\cR}(S,Set)$}), to each one adding $E_A = A, E_C = C$ with $\sigma_A = 1, \sigma_C = 0, m_A = 2, m_C = |V|$ (similar to above).
\end{itemize}
Compute the remaining terms in the expectation of $\E{I^\uparrow_S \mid Set}$ analogously.
\end{mybox}

For an edge bundle $\bbe$ and a set $A \subseteq E$ we use the shorthand $\bbe(A)$ to denote the set of edges in $A$ and $\bbe$. We now define $I^\rightarrow(S)$ for a near-cycle cut $S$.  There are three cases:
\begin{itemize} 
\item \textbf{Case 1:} The parent $\hat{S}$ of $S$ is a polygon cut. Then define
$$I^\rightarrow(S) := (1+\eps_\eta)\beta \left( \max\{x(A^\rightarrow),x(B^\rightarrow)\} + x(C^\rightarrow)\right) \cdot \I{\cR_{\hat{S}} = 1, S \text{ not happy}}$$

\item \textbf{Case 2:} The parent $u$ of $S$ is a degree cut with degree partition $A',B',C'$ and has a pair $\bbe = (v,S), \bbf= (w,S)$ edges with \hyperlink{type}{{\tt type}$(\bbe,S)$}$=$ \hyperlink{type}{{\tt type}$(\bbf,S)$}$=\text{2-2-2}$. WLOG, let  $x_{\bbe(B')}, x_{\bbf(A')} \le \epsilon_{1/2}$. Then define:
$$I^\rightarrow(S) := (1+\eps_\eta) \frac{\tau}{2}  \max\{x_{\bbe(A')}, x_{\bbf(B')}\}\left(\I{\cR_{\{\bbe,\bbf\}}}  + \I{\cR_{\bbe,v}} + \I{\cR_{\bbf,w}}\right)$$
$$ + (1+\eps_\eta) \sum_{g \in \delta^\rightarrow(S) \smallsetminus \bbe(A') \smallsetminus \bbf(B')} r_g \cdot \I{S \text{ not happy}}.$$
\item \textbf{Case 3:} Otherwise the parent $u$ of $S$ is a degree cut with no pair of 2-2-2 edges. Define:
 $$I^\rightarrow(S) := (1+\eps_\eta) \sum_{g \in \delta^\rightarrow(S)} \E{r_g \cdot \I{S \text{ not happy}} \mid Set}.$$
\end{itemize}

\begin{mybox}{gray}{\hypertarget{IrightSet}{$\mathbb{E}_{I^\rightarrow}(S,Set)$}}
Assume $S$ is a near-cycle cut. This function computes $\E{I^\rightarrow(S) \mid Set}$ for the above three cases as follows (using the notation from above):
\begin{itemize} 
\item \textbf{Case 1:} Return
$$(1+\eps_\eta)\beta \left( \max\{x(A^\rightarrow),x(B^\rightarrow)\} + x(C^\rightarrow)\right) \cdot \E{\cR_{\hat{S}} \cdot \I{S \text{ not happy}} \mid Set}$$
Here to apply \cref{lem:master} we sum over the events for $\cR_{\hat{S}}$ as in \eqref{eq:rpolygon} (used by \hyperlink{RpolySet}{$\mathbb{E}_{\cR}(S,Set)$}), to each one adding $E_A=A, E_B = B, E_C=C$ and $\sigma_A = 1, \sigma_B = 1, \sigma_C = 0, m_A = 2, m_B=2, m_C=|V|$.\footnote{Note that the sets $A,B,C$ written here are from the near-cycle partition of $S$, however the event $\cR_{\hat{S}}$ uses the near-cycle partition of $\hat{S}$.}

\item \textbf{Case 2:} Return
$$(1+\eps_\eta) \frac{\tau}{2}  \max\{x_{\bbe(A')}, x_{\bbf(B')}\}\left(\P{\cR_{\{\bbe,\bbf\}}\mid Set}  + \P{\cR_{\bbe,v} \mid Set} + \P{\cR_{\bbf,w} \mid Set}\right)$$
$$ + (1+\eps_\eta) \sum_{g \in \delta^\rightarrow(S) \smallsetminus \bbe(A') \smallsetminus \bbf(B')} \E{r_g \cdot \I{S \text{ not happy}} \mid Set}.$$
Recall $\E{\cR_{\{\bbe,\bbf\}} \mid Set}$ can be computed by \hyperlink{ReSet}{$\mathbb{E}_{\cR}(\bbe,S,Set)$}. To calculate $\E{r_g \cdot \I{S \text{ not happy}} \mid Set}$ we use techniques similar to \hyperlink{IupSet}{$\mathbb{E}_{I^\uparrow}(S,Set)$}. 
\item \textbf{Case 3:} Return
 $$(1+\eps_\eta) \sum_{g \in \delta^\rightarrow(S)} \E{r_g \cdot \I{S \text{ not happy}} \mid Set}.$$
\end{itemize}
\end{mybox}

\begin{mybox}{gray}{\hypertarget{ISSet}{$\mathbb{E}_{I}(S,Set)$}}
	Return $\hyperlink{IupSet}{\mathbb{E}_{I^\uparrow}(S,Set)}+\hyperlink{IrightSet}{\mathbb{E}_{I^\rightarrow}(S,Set)}$.
\end{mybox}

\subsubsection{Increases for top edges}

For each top edge bundle $\bbe= (u,v)$, using the values $m_{\bbe,u},m_{\bbe,v}$ from \cref{subsec:matching}, define for a tree $T$:
\begin{equation}
I_{\bbe,u} := \sum_{g \in \delta^\uparrow(u)} r_g \cdot \I{\delta(u)_T \text{ is odd}}\cdot\frac{ \hypertarget{preprocessing6}{m_{\bbe,u}}}{ \sum_{\bbf\in\delta^\rightarrow(u)} \hypertarget{preprocessing6}{m_{\bbf,u}}}
\end{equation}
and define $I_{\bbe,v}$ analogously. We then let 
$$I_\bbe := I_{\bbe,u} + I_{\bbe,v}.$$

\begin{mybox}{gray}{\hypertarget{IeSet}{$\mathbb{E}_{I}(\bbe,Set)$}}
	Let $\bbe=(u,v)$. Compute:
\begin{equation}
\E{I_{\bbe,u} \mid Set} := \sum_{g \in \delta^\uparrow(u)} \E{r_g \cdot \I{\delta(u)_T \text{ is odd}}\mid Set}\cdot\frac{ \hyperlink{preprocessing6}{m_{\bbe,u}}}{ \sum_{\bbf\in\delta^\rightarrow(u)} \hyperlink{preprocessing6}{m_{\bbf,u}}}
\end{equation}
and compute $\E{I_{\bbe,v} \mid Set}$ analogously, using the values $m_{\bbe,u}$ computed previously in Step 6. To compute the above, we apply \cref{lem:master} for sets $E_i$ coming from $\hyperlink{littlereset}{\mathbb{E}_{r}(g,Set)}$ and add an extra set $E_u = \delta(u)$ and coordinates $\sigma_u = 1, m_u = 2$. 

Return 
$$\E{I_\bbe \mid Set} := \E{I_{\bbe,u} + I_{\bbe,v} \mid Set}.$$
\end{mybox}

\subsection{Computation of $\E{c(s) \mid Set}$}

First we define $s^{\cH}$:
\begin{equation}\label{eq:sH}
		s^{\cH}_e := -r_e + \begin{cases} I_\bbf \frac{x_e}{x_{\bbf}} & \text{ if $e\in \bbf$ for a top edge bundle $\bbf$,}\\
	I_S  x_e & \text{ if $\p(e)=S$ for a polygon cut $S\in\cH$, i.e. $e$ is a bottom edge.}
 \end{cases}
\end{equation}


Finally, we construct $s$. Note that $s^{\cH}_e = 0$ with probability 1 for a bad edge bundle $e$. Therefore in \cite{KKO21b} a second slack vector was defined to allow bad edges to reduce. In particular, let $E_g$ be the set of good edges and let $E_b:=E\smallsetminus E_g$ be the set of bad edges. Note all edges in $\delta(\{u_0,v_0\})$ are bad edges as they are not edge bundles in the hierarchy. Define the vector $s^{bad}:E\cup \{e_0\}\to\R$ as follows: 
\begin{equation}\label{eq:tildesdef}s^{bad}_e \gets \begin{cases}\infty & \text{if } e=e_0\\
-x_e(4\decrease/5)(1-2\eta) & \text{if } e\in E_b,\\
x_e(4\decrease/3) & \text{otherwise.}
\end{cases}
\end{equation}
Finally, where $\gamma = \frac{15}{32}\eps_P$, let $s= \gamma s^{bad} + (1-\gamma)s^\cH$. This is now exactly the vector $s$ from Theorem 6.1 of \cite{KKO21b}.

\begin{mybox}{yellow}{\hypertarget{cs}{$\mathbb{E}_{c(s)}(e,Set)$}}\label{alg:c(s)}
	We have $\E{c(s_e) \mid Set} = \gamma c(s^{bad}_e) + (1-\gamma)\E{s^\cH_e \mid Set}$. $c(s^{bad}_e)$ is a constant which can be computed by \eqref{eq:tildesdef}.\\
	
	Thus, it is sufficient to compute $\E{c(s^{\cH}_e) \mid Set}$. From \eqref{eq:sH}, we just need $\E{r_e \mid Set}$, computed by \hyperlink{littlereset}{$\mathbb{E}_{r}(e,Set)$}, and $\E{I_\bbf \mid Set}$ if $e \in \bbf$ is a top edge (computed by \hyperlink{IeSet}{$\mathbb{E}_{I}(\bbf,Set)$}) and $\E{I_S \mid Set}$ if it is a bottom edge with near-cycle parent $S$ (computed by \hyperlink{ISSet}{$\mathbb{E}_{I}(S,Set)$}).
\end{mybox}
This concludes the proof of (2) of \cref{lem:compute-s}.

%% file: tsp.bib
@article{HNR21,
  author    = {Arash Haddadan and
               Alantha Newman and
               R. Ravi},
  title     = {Shorter tours and longer detours: uniform covers and a bit beyond},
  journal   = {Math. Program.},
  volume    = {185},
  number    = {1-2},
  pages     = {245--273},
  year      = {2021},
  url       = {https://doi.org/10.1007/s10107-019-01426-8},
  doi       = {10.1007/s10107-019-01426-8},
  bibsource = {dblp computer science bibliography, https://dblp.org}
}

@article{GLLM21,
  author    = {Anupam Gupta and
               Euiwoong Lee and
               Jason Li and
               Marcin Mucha and
               Heather Newman and
               Sherry Sarkar},
  title     = {Matroid-Based {TSP} Rounding for Half-Integral Solutions},
  journal   = {CoRR},
  volume    = {abs/2111.09290},
  year      = {2021},
  url       = {https://arxiv.org/abs/2111.09290},
  eprinttype = {arXiv},
  eprint    = {2111.09290},
  bibsource = {dblp computer science bibliography, https://dblp.org}
}

@article{KLS15,
title = "New inapproximability bounds for TSP",
journal = "Journal of Computer and System Sciences",
volume = "81",
number = "8",
pages = "1665 - 1677",
year = "2015",
issn = "0022-0000",
author = "Marek Karpinski and Michael Lampis and Richard Schmied",
}

@article{Ser78,
title={O nekotorykh ekstremal’nykh obkhodakh v grafakh}, 
journal={Upravlyaemye sistemy},
volume={17}, 
pages={76--79},
year={1978},
author={Serdyukov, A. I.},
url={http://nas1.math.nsc.ru/aim/journals/us/us17/us17_007.pdf},
}

@inproceedings{TVZ20,
  author    = {Vera Traub and
               Jens Vygen and
               Rico Zenklusen},
  editor    = {Konstantin Makarychev and
               Yury Makarychev and
               Madhur Tulsiani and
               Gautam Kamath and
               Julia Chuzhoy},
  title     = {Reducing path {TSP} to {TSP}},
  booktitle = {STOC},
  pages     = {14--27},
  publisher = {ACM},
  year      = {2020},
}

@inproceedings{HN19,
  author    = {Arash Haddadan and
               Alantha Newman},
  editor    = {Michael A. Bender and
               Ola Svensson and
               Grzegorz Herman},
  title     = {Towards Improving Christofides Algorithm for Half-Integer {TSP}},
  booktitle = {ESA},
  series    = {LIPIcs},
  volume    = {144},
  pages     = {56:1--56:12},
  publisher = {Schloss Dagstuhl - Leibniz-Zentrum f{\"{u}}r Informatik},
  year      = {2019},
}

@inproceedings{OSS11,
 author = {{Oveis Gharan}, Shayan and Saberi, Amin and Singh, Mohit},
 title = {A Randomized Rounding Approach to the Traveling Salesman Problem},
 booktitle = {FOCS},
 year = {2011},
 isbn = {978-0-7695-4571-4},
 pages = {550--559},
 numpages = {10},
  publisher = {IEEE Computer Society},
 }

@article{DFJ59,
	Author = {G.B. Dantzig and D.R. Fulkerson and S. Johnson},
	Journal = {OR},
	Pages = {58-66},
	Title = {On a Linear Programming Combinatorial Approach to the Traveling Salesman Problem},
	Volume = {7},
	Year = {1959}}

@inproceedings{Muc12,
	Author = {Mucha, M},
	Booktitle = {STACS},
	Pages = {30--41},
	Title = {$\frac{13}{9}$-approximation for graphic TSP.},
	Year = {2012}}

@unpublished{SV12,
	Author = {Andr{\'a}s Seb{\"o} and Jens Vygen},
	Note = {CoRR abs/1201.1870},
	Title = {Shorter Tours by Nicer Ears:},
	Year = {2012}}

@inproceedings{MS11,
	Author = {Moemke, Tobias and Svensson, Ola},
	Booktitle = {FOCS},
	Pages = {560--569},
	Title = {Approximating Graphic TSP by Matchings},
	Year = {2011}}

@article{NNI94,
	Author = {Hiroshi Nagamochi and Kazuhiro Nishimura and Toshihide Ibaraki},
	Journal = {SIAM Journal on Discrete Mathematics},
	Pages = {469--481},
	Title = {Computing All Small Cuts in an Undirected Network},
	Volume = {10},
	Year = {1994}}

@article{EJ73,
	Author = {Edmonds, Jack and Johnson, Ellis L.},
	Journal = {Mathematical Programming},
	Number = {1},
	Pages = {88--124},
	Publisher = {Springer Berlin / Heidelberg},
	Title = {Matching, Euler tours and the Chinese postman},
	Volume = {5},
	Year = {1973}}

@article{BG08,
	Author = {Andr{\'a}s A. Bencz{\'u}r and Michel X. Goemans},
	Journal = {Building Bridges: Between Mathematics and Computer Science, M. Groetschel and G.O.H. Katona, Eds., Bolyai Society Mathematical Studies},
	Pages = {103--135},
	Title = {Deformable Polygon Representation and Near-Mincuts},
	Volume = {19},
	Year = {2008}}

@inproceedings{Ben95,
	Author = {Andr{\'a}s A. Bencz{\'u}r},
	Booktitle = {FOCS},
	Pages = {92-102},
	Title = {A Representation of Cuts within 6/5 Times the Edge Connectivity with Applications},
	Year = {1995}}

@inproceedings{Edm70,
	Address = {New York, NY, USA},
	Author = {Edmonds, Jack},
	Booktitle = {Combinatorial Structures and Their Applications},
	Pages = {69--87},
	Publisher = {Gordon and Breach},
	Title = {Submodular functions, matroids and certain polyhedra},
	Year = {1970}}

@book{ABCC07,
	Address = {Princeton, NJ, USA},
	Author = {Applegate, David L. and Bixby, Robert E. and Chvatal, Vasek and Cook, William J.},
	Publisher = {Princeton University Press},
	Title = {The Traveling Salesman Problem: A Computational Study (Princeton Series in Applied Mathematics)},
	Year = {2007}}

@inproceedings{AGMOS10,
	Author = {Arash Asadpour and Michel X. Goemans and Aleksander Madry and Shayan {Oveis Gharan} and Amin Saberi},
	Booktitle = {SODA},
	Pages = {379-389},
	Title = {An O(log n/ log log n)-approximation Algorithm for the Asymmetric Traveling Salesman Problem},
	Year = {2010}}

@phdthesis{Ben97,
	Author = {Andras A. Bencz{\'u}r},
	Order_No = {AAI0598649},
	Publisher = {Massachusetts Institute of Technology},
	School = {MIT},
	Title = {Cut structures and randomized algorithms in edge-connectivity problems},
	Year = {1997}}

@article{BG93,
	author = {Goemans, Michel and Bertsimas, Dimitris},
	year = {1993},
	month = {06},
	pages = {},
	title = {Survivable Network, Linear Programming Relaxations and the Parsimonious Property},
	volume = {60},
	journal = {Math Program},
	doi = {10.1007/BF01580607}
}

@techreport{Chr76,
	Address = {Pittsburgh, PA},
	Author = {Nicos Christofides},
	Institution = {Graduate School of Industrial Administration, Carnegie-Mellon University},
	Number = {388},
	Title = {Worst Case Analysis of a New Heuristic for the Traveling Salesman Problem},
	Type = {Report},
	Year = {1976}}

@article{HK70,
	Author = {M. Held and R.M. Karp},
	Journal = {Operations Research},
	Pages = {1138--1162},
	Title = {The traveling salesman problem and minimum spanning trees},
	Volume = {18},
	Year = {1970}}

@inproceedings{KKO21a,
      title={A (Slightly) Improved Approximation Algorithm for Metric TSP}, 
      author={Anna R. Karlin and Nathan Klein and Shayan {Oveis Gharan}},
      year={2021},
      booktitle={STOC},
      publisher={ACM},
}

@inproceedings{KKO21b,
      title={A (Slightly) Improved Bound on the Integrality Gap of the Subtour LP for TSP}, 
      author={Anna Karlin and Nathan Klein and Shayan {Oveis Gharan}},
      year={2022},
      booktitle={FOCS},
      pages = {844-855},
      publisher = {IEEE Computer Society},
}

@inproceedings{KKO20,
  author    = {Anna R. Karlin and
               Nathan Klein and
               Shayan {Oveis Gharan}},
  editor    = {Konstantin Makarychev and
               Yury Makarychev and
               Madhur Tulsiani and
               Gautam Kamath and
               Julia Chuzhoy},
  title     = {An improved approximation algorithm for {TSP} in the half integral
               case},
  booktitle = {STOC},
  pages     = {28--39},
  publisher = {{ACM}},
  year      = {2020},
}
